\documentclass[10pt,journal,compsoc]{IEEEtran}
\ifCLASSOPTIONcompsoc
  \usepackage[nocompress]{cite}
\else
  \usepackage{cite}
\fi
\ifCLASSINFOpdf
\else
\fi

\usepackage{amssymb,amsmath,amsthm}
\usepackage[utf8x]{inputenc}
\usepackage{epsfig,listings}
\usepackage{url,cite}
\usepackage{xspace}
\usepackage{graphicx,xcolor}
\usepackage{algorithm,algpseudocode,lineno}

\usepackage{algorithm}
\usepackage{algpseudocode}
\algblock{Input}{EndInput}
\algnotext{EndInput}
\algblock{Output}{EndOutput}
\algnotext{EndOutput}

\usepackage{enumitem}

\newtheorem{corollary}{Corollary}

\newtheorem{theorem}{Theorem}

\newtheorem{remark}{Remark}

\newcommand\calL{{\mathcal L}}
\newcommand\calI{{\mathcal I}}
\newcommand\calD{{\mathcal D}}
\newcommand{\calZ}{\mathcal Z}
\newcommand{\calY}{\mathcal Y}
\newcommand{\calG}{\mathcal G}

\newcommand{\calF}{\mathcal F}
\newcommand{\calN}{\mathcal N}
\newcommand\inc{\mathrm{In}}
\newcommand\out{\mathrm{Out}}

\newcommand\pred{\mathrm{Pred}}
\newcommand\lo{\mathrm{Low}}
\newcommand\paf{\mathrm{Path}}
\newcommand\dep{\mathrm{Level}}

\newcommand{\lp}{\ensuremath{\left(}}
\newcommand{\rp}{\ensuremath{\right)}}

\newcommand{\lc}{\ensuremath{\left\{}}
\newcommand{\rc}{\ensuremath{\right\}}}
\newcommand{\abs}[1]{\left\vert#1\right\vert}

\def\Reals{\mathbb{R}}

\def\be{\begin{equation}}
\def\ee{\end{equation}}
\def\ben{\begin{equation*}}
\def\een{\end{equation*}}
\def\bearn{\begin{eqnarray*}}
\def\eearn{\end{eqnarray*}}
\def\bear{\begin{eqnarray}}
\def\eear{\end{eqnarray}}
\def\barr{\begin{array}}
\def\earr{\end{array}}

\newcommand{\mst} {\mathrm{\;  such \;  that \; }}

\newcommand{\nin}{\not\in}

\newcommand{\eqdef}{\ensuremath{\overset{\scriptsize\textit{def}}{=}}}

\newcommand{\cref}[1]{Chapter~\ref{#1}}

\newcommand{\thref}[1]{Theorem~\ref{#1}}

\newcommand\hd{\textit{(H1)}\xspace}

\newcommand\synctfa{SyncTFA\xspace}

\newcommand\asynctfa{AsyncTFA\xspace}
\newcommand\asyncalttfa{AltTFA\xspace}

\newcommand\fptfa{FPTFA\xspace}
\newcommand\tfa{TFA\xspace}

\newcommand\ie{\mbox{i.e.}\xspace}

\usepackage{tikz}
\usepackage{figlatex}
\usepackage{url}
\usepackage{array}
\usetikzlibrary{automata,arrows,positioning,calc,math,arrows.meta}
\usetikzlibrary{shapes,positioning}
\usetikzlibrary{fit,intersections,decorations.pathreplacing,matrix}

\tikzset{
    ncbar angle/.initial=90,
    ncbar/.style={
        to path=(\tikztostart)
        -- ($(\tikztostart)!#1!\pgfkeysvalueof{/tikz/ncbar angle}:(\tikztotarget)$)
        -- ($(\tikztotarget)!($(\tikztostart)!#1!\pgfkeysvalueof{/tikz/ncbar angle}:(\tikztotarget)$)!\pgfkeysvalueof{/tikz/ncbar angle}:(\tikztostart)$)
        -- (\tikztotarget)
    },
    thick,
    ncbar/.default=0.1cm,
}
\tikzset{square left brace/.style={ncbar=0.1cm}}
\tikzset{square right brace/.style={ncbar=-0.1cm}}
\tikzset{round left paren/.style={ncbar=0.1cm,out=120,in=-120}}
\tikzset{round right paren/.style={ncbar=0.1cm,out=60,in=-60}}
\newcommand\dgreen{green!50!black}

\usepackage{empheq}

\begin{document}
\title{Equivalent Versions of Total Flow Analysis}
%
%
%
%

\author{Stéphan~Plassart and Jean-Yves~Le Boudec
\IEEEcompsocitemizethanks{\IEEEcompsocthanksitem \'Ecole Polytechnique F\'ed\'erale de Lausanne (EPFL), Switzerland.\protect\\
E-mail: $\{$firstname.lastname$\}$@epfl.ch}
\thanks{This work was supported by Huawei Technologies Co., Ltd. in the framework of the project Large Scale Deterministic Network.}}

%
%

\markboth{Journal of \LaTeX\ Class Files,~Vol.~14, No.~8, August~2015}%
{Shell \MakeLowercase{\textit{et al.}}: Bare Demo of IEEEtran.cls for Computer Society Journals}

\IEEEtitleabstractindextext{%

\begin{abstract}
Total Flow Analysis (TFA) is a method for conducting the worst-case analysis of time-sensitive networks that are without cyclic dependencies.  
In networks with cyclic dependencies, Fixed-Point TFA introduces artificial cuts, analyses the resulting cycle-free network with TFA, and iterates on it.  
If it converges, it does provide valid performance bounds.  
We show that the choice of the specific cuts used by Fixed-Point TFA does not affect its convergence or the obtained performance bounds, and that it can be replaced by an alternative algorithm that does not use any cuts at all, while still applying to cyclic dependencies.
\end{abstract}

\begin{IEEEkeywords}
Delay  bound,  service  curve,  network  calculus, FIFO system, deterministic  networks, Total Flow Analysis (TFA)
\end{IEEEkeywords}}

\maketitle

\IEEEdisplaynontitleabstractindextext

\section{Introduction}
\label{sec:intro}

In time-sensitive networks,  obtaining deterministic bounds is required on worst case delay, delay-jitter,  and backlog. Total Flow Analysis (TFA) is a method for conducting worst-case analysis is such settings for per-class networks~\cite{bouillardbook}.  
It is implemented in industrial software (WoPANets~\cite{wopanet}) and can consider several important features such as the effect of packetization, regulators, and line shaping~\cite{mifdaoui2017beyond}.  
In a per-class network, the traffic of a given class is isolated from other classes by using mechanisms 
such as Deficit Round-Robin \cite{boyer2012deficit} or the Credit-Based Shaper\cite{daigmorte2018modelling}; 
inside a class, packets of all flows are handled in a FIFO manner.  
One of the main issues is that the burstiness of a flow increases as the flow travels through the network.  
If there is no cyclic dependency, i.e., if the graph induced by flow paths has no cycle, TFA first analyses edge-nodes by using network calculus~\cite[Section 1.4]{LebBook},  
then computes hop-by-hop the propagated burstiness.  
Propagated burstiness is derived from bounds on delay-jitters between a source and a point inside the network; 
such a bound is the sum of the delay-jitter bounds at the nodes on the path of a flow.

Computing performance bounds when there are cyclic dependencies is more challenging, 
as it is generally not known under which conditions deterministic bounds exist~\cite{UnboundedAndrews09}.  
For such cases, the most recent version of TFA, Fixed-Point TFA (\fptfa~\cite{thomas2019cyclic}) introduces artificial network \emph{cuts}; 
the resulting cycle-free network is analysed, and the output is fed back into the analysis. 
If the scheme converges, which is assumed to occur when the utilization is not too large, 
then \cite{thomas2019cyclic} shows that the resulting fixpoint provides valid performance bounds.  
\cite{thomas2019cyclic} shows that finding the minimal number of cuts to realize on a cyclic network is the most time consuming part of their algorithm, which raises the following question: 
Do the results of the worst-case analysis performed by \fptfa depend on the specific cuts chosen by the algorithm~? 
Are some cuts better than others~? 
Is there benefit in computing minimal cuts~?

We show that the answer to all of these questions is negative: cuts do not matter. 
We obtain this by investigating alternative formulations of TFA that do not make cuts, while still applying to networks with cyclic dependencies.
All formulations presented in this article are based on the same principle of TFA,
but they differ by the instant and the nodes at which the burstiness and the delay-jitter bounds are updated.
We distinguish two types of algorithms:
\begin{itemize}
\item synchronous: at every iteration, delay-jitter and burstiness bounds are updated at all nodes, based on the values of delay-jitter and burstiness bounds at the previous iteration;
\item asynchronous: at every iteration, a set of nodes is visited, delay-jitter and output burstiness bounds are updated based on the burstiness bounds computed in previous iterations; 
nodes are visited according to some arbitrary scheme.
\end{itemize} 
The synchronous algorithm is of theoretical use, it serves to show the equivalence of all schemes. 
The asynchronous algorithm is practical, as it combines a flexible choice of the sets of nodes visited at every iteration with the use of the most recent information. 
The alternating TFA algorithm, which is of particular interest in symmetric networks, is a special case of the  asynchronous algorithm.

The contributions of this paper are as follows:
\begin{itemize}
	\item We introduce versions of TFA that do not make cuts and apply to networks with or without cyclic dependencies. 
	We show that all of these versions are equivalent; furthermore they are equivalent to TFA in networks without cyclic dependencies and to \fptfa in networks with cyclic dependencies.  
	All algorithms converge or they all diverge; and if they all converge, they all give the same performance bounds. 
	\item It follows that the behaviour and the result obtained by \fptfa do not depend on the chosen cut. 
	\item We prove that all algorithms are correct, i.e., if they converge, they do provide valid delay-jitter bounds. 
	The proof does not require any assumption on propagation times, unlike the original proof of \fptfa.
	\item To obtain these results, we provide a variant of the Knaster-Tarski fixpoint theorem and a new fixpoint theorem for concave functions. 
\end{itemize}

In Section~\ref{sec:notation}, we introduce the system model and, in Section~\ref{sec:SoA}, we describe TFA and \fptfa in a compact form suitable for analysis. 
In Section~\ref{sec:sync-tfa},  we describe Synchronous TFA,  and establish its validity (Theorem~\ref{th:lowGbounded}) and its equivalence with TFA/~\fptfa (Theorem~\ref{th:fptfa}). 
In Section~\ref{sec:async-tfa}, we do the same for Asynchronous TFA (Theorem~\ref{th:theo-async}) and introduce Alternating TFA as a special case.  
In Section~\ref{sec:fixpointsMaxElem},  we introduce a variant of the Knaster-Tarski fixpoint theorem (Theorem~\ref{theo-tarski}) and a new fixpoint theorem for concave functions (Theorem~\ref{theo-fpconc}); 
both are at the heart of the theoretical analysis performed in the previous sections.
In Section~\ref{sec:simu}, we give a numerical illustration to a ring network.

\section{System Model}
\label{sec:notation}
\subsection{Time-Sensitive Packet-Switched Network}
We consider a packet-switched network.
Each device in the network is composed of input ports, output ports,  and a switch fabric,  for example in Figure~\ref{fig:net}.
Each input port contains a packetizer that releases 
the data when an entering packet has been entirely received. 
The packet is then transmitted through the switch fabric to the scheduler of one specific output port (as indicated by the routing table);
then,  it is serialized on the output line at the transmission rate of the line.
The output port schedulers that separate traffic classes and are first-in-first-out (FIFO) inside a class.
Each scheduler has a minimum service curve that is assumed to be a rate-latency service curve \cite{LebBook}.
This models the service isolation provided to a traffic class;
for example, rate-latency service curves are given in \cite{boyer2012deficit} for Deficit Round-Robin and \cite{daigmorte2018modelling} for the Credit-Based Shaper.  
A rate-latency service curve has two parameters: a rate that captures the service rate guaranteed to the class; 
and a latency that captures an additional delay that has to be considered when computing delay bounds.

In the remainder of the paper,  we focus on one of these traffic classes that is assumed to receive a deterministic service \cite{802.1Q-2018}.  
This means that every flow in the class (which can be unicast or multicast) has a fixed path and is constrained at the source.  
Specifically, we assume that flow $f$ is constrained at its source by an arrival curve $\alpha_f$ of the form:
$$\alpha_f(t) = r_f t +b_f,$$
\ie,  a ``leaky bucket" arrival curve with rate $r_f>0$ and burstiness $b_f>0$.
The arrival curve constraint means that over any time interval of any duration $t$, flow $f$ should not generate more than $\alpha_f(t)$ bits.  
Let $L_{max} > 0$ and $L_{min}$ be,  respectively,  the maximum and minimum packet size over all flows. 

The deterministic service means that bounds on delay,  delay-jitter (defined as the difference between worst-case and best-case delays) and backlog are computed at every node, using the source arrival curves and the rate-latency service curves.  
The challenge with such computations is that the burtiness of a flow increases after crossing a node and the resulting ``propagated burstiness'' needs to be estimated. 
We assume that local stability holds at every node $i$, \ie, the sum of the rates $r_f$ for all flows that cross node~$i$ is less than or equal to the rate of the rate-latency service curve of node~$i$.
This is a necessary condition for the existence of finite performance bounds (but not sufficient \cite{UnboundedAndrews09}).

\subsection{Cyclic Dependencies and the Graph Induced by Flows}
\begin{figure}
  \centering
  \scalebox{0.75}{\begin{tikzpicture}[->, auto, semithick, draw=black, >=stealth]

\tikzset{
	source/.style = {star,star points=7,star point ratio=0.8, draw,thick},
	switch/.style = {rectangle, draw, thick, minimum width=0.7cm, minimum height = 2cm},
	switchleg/.style = {rectangle, draw, thick, minimum width=0.2cm, minimum height = 0.5cm},
    pre/.style =    {<-, semithick},
    post/.style =   {->, semithick}
}

\node[source] (S1) at (-0.3,5.2) {\footnotesize S1};
\node[source] (S2) at (0.2,2.53) {\footnotesize S2};
\node[source] (S3) at (10.3,2.1) {\footnotesize S3};
\node[source] (S4) at (10.3,1.1) {\footnotesize S4};

\node[switch] (s1) at (1,1) {$\times$};
\node[switch] (s2) at (3,1) {$\times$};
\node[switch] (s3) at (5,1) {$\times$};
\node[switch] (s4) at (7,1) {$\times$};
\node[switch] (s5) at (9,1) {$\times$};

\node[switch] (s6) at (1,4) {$\times$};
\node[switch] (s7) at (3,4) {$\times$};
\node[switch] (s8) at (5,4) {$\times$};
\node[switch] (s9) at (7,4) {$\times$};
\node[switch] (s10) at (9,4) {$\times$};

\draw [-,black, thick] (0.52,0) to [square left brace] (0.52,0.5); \draw[fill opacity=0] (0.52,0) node[below,fill opacity=1] {\scriptsize os2 \,\,};
\draw [-,black, thick] (0.52,0.7) to [square left brace] (0.52,1.2); \draw[fill opacity=0] (0.52,1.2) node[left,fill opacity=1] {\scriptsize O1};
\draw [-,black, thick] (0.52,1.4) to [square left brace] (0.52,1.9); \draw[fill opacity=0] (0.52,1.9) node[above,fill opacity=1] {\scriptsize os3 \,\,};

\draw [-,black, thick] (1.58,0) to [square left brace] (1.58,1.9);

\draw [-,black, thick] (2.52,0) to [square left brace] (2.52,1.9); \draw[fill opacity=0] (2.52,1.9) node[above,fill opacity=1] {\scriptsize O2 \,\,};

\draw [-,black, thick] (3.58,0) to [square left brace] (3.58,1.9);

\draw [-,black, thick] (4.52,0) to [square left brace] (4.52,1.2); \draw[fill opacity=0] (4.52,0) node[below,fill opacity=1] {\scriptsize O3 \,\,};
\draw [-,black, thick] (4.52,1.4) to [square left brace] (4.52,1.9); \draw[fill opacity=0] (4.52,1.9) node[above,fill opacity=1] {\scriptsize O11 \,\,};

\draw [-,black, thick] (5.58,0) to [square left brace] (5.58,1.2);
\draw [-,black, thick] (5.58,1.4) to [square left brace] (5.58,1.9);

\draw [-,black, thick] (6.52,0) to [square left brace] (6.52,1.9); \draw[fill opacity=0] (6.52,1.9) node[above,fill opacity=1] {\scriptsize O4 \,\,};

\draw [-,black, thick] (7.58,0) to [square left brace] (7.58,1.9);

\draw [-,black, thick] (8.52,0) to [square left brace] (8.52,1.9); \draw[fill opacity=0] (8.52,1.9) node[above,fill opacity=1] {\scriptsize O5 \,\,};

\draw [-,black, thick] (9.58,0) to [square left brace] (9.58,0.5);
\draw [-,black, thick] (9.58,0.7) to [square left brace] (9.58,1.2);
\draw [-,black, thick] (9.58,1.4) to [square left brace] (9.58,1.9);

\draw [-,black, thick] (0.4,3) to [square right brace] (0.4,3.5);
\draw [-,black, thick] (0.4,3.7) to [square right brace] (0.4,4.2);
\draw [-,black, thick] (0.4,4.4) to [square right brace] (0.4,4.9);

\draw [-,black, thick] (1.45,3) to [square right brace] (1.45,4.9); \draw[fill opacity=0] (1.45,4.9) node[above,fill opacity=1] { \,\, \scriptsize O6};

\draw [-,black, thick] (2.4,3) to [square right brace] (2.4,4.9);

\draw [-,black, thick] (3.45,3) to [square right brace] (3.45,4.9); \draw[fill opacity=0] (3.45,4.9) node[above,fill opacity=1] {\,\, \scriptsize O7};

\draw [-,black, thick] (4.4,3) to [square right brace] (4.4,3.5);
\draw [-,black, thick] (4.4,3.7) to [square right brace] (4.4,4.2);
\draw [-,black, thick] (4.4,4.4) to [square right brace] (4.4,4.9);

\draw [-,black, thick] (5.45,3) to [square right brace] (5.45,3.5);\draw[fill opacity=0] (5.45,3)  node[below,fill opacity=1] { \,\, \scriptsize O12};
\draw [-,black, thick] (5.45,3.7) to [square right brace] (5.45,4.9); \draw[fill opacity=0] (5.45,4.9) node[above,fill opacity=1] { \,\, \scriptsize O8};

\draw [-,black, thick] (6.4,3) to [square right brace] (6.4,4.9);

\draw [-,black, thick] (7.45,3) to [square right brace] (7.45,4.9); \draw[fill opacity=0] (7.45,4.9) node[above ,fill opacity=1] {\,\, \scriptsize O9};

\draw [-,black, thick] (8.4,3) to [square right brace] (8.4,4.9);

\draw [-,black, thick] (9.45,3) to [square right brace] (9.45,3.5); \draw[fill opacity=0] (9.45,3) node[below,fill opacity=1] {\,\, \scriptsize os4};
\draw [-,black, thick] (9.45,3.7) to [square right brace] (9.45,4.2); \draw[fill opacity=0] (9.45,3.7) node[right,fill opacity=1] {\,\scriptsize os1};
\draw [-,black, thick] (9.45,4.4) to [square right brace] (9.45,4.9); \draw[fill opacity=0] (9.45,4.9) node[above,fill opacity=1] {\,\, \scriptsize O10};

\draw[blue,ultra thick,dashed] (S1) -- (-0.1,4.7) -- (10.8,4.7) -- (10.8,0.3) node[midway, right, blue] {$f_{bl}$} -- (0,0.3);
\draw[red,ultra thick] (S4) -- (9.9,0.8) -- (-0.3,0.8) -- (-0.3,4.1) node[midway, left, red] {$f_r$} -- (10.3,4.1);
\draw[\dgreen,ultra thick,dash dot] (S2) -- (0.2,3.3) -- (0.8,3.3) -- (1.2,3.9) -- (4.8,3.9) -- (5.2,3.3) -- (5.6,3.3) -- (6,2.5) -- (5.6,1.6) -- (5.2,1.6)  -- (4.8,1) -- (1.3,1)-- (0.9,1.6) -- (0,1.6);
\draw[color={rgb:red,4;green,2;yellow,1},ultra thick,dotted] (S3) -- (9.8,1.6) -- (9.2,1.6) -- (8.8,1.1) -- (5.2,1.1) -- (4.8,1.7) -- (4.2,1.7) -- (3.8,2.5) -- (4.2,3.2) -- (4.8,3.2) -- (5.3,3.9) -- (8.8,3.9) -- (9.2,3.3) -- (10.3,3.3);
\draw[fill opacity=0] (6,2.5) node[fill opacity=1,right, \dgreen] {$f_g$};
\draw[fill opacity=0] (3.8,2.5) node[fill opacity=1,left, color={rgb:red,4;green,2;blue,1}] {$f_{br}$};
\node[source] (legS) at (0,-0.7) {};
\node[right = 0.2cm of legS] {Flow source};
\node[thick] (legP) at (3,-0.7) {]};
\node[right = 0.2cm of legP] {Ports};
\node[switchleg] (legSw) at (5.5,-0.7) {$\times$};
\node[right = 0.2cm of legSw] {Switch fabric};
\draw (-0.2,-1.2) -- (0.5,-1.2) node[right] {Flow};
\node[circle,draw] (Cc) at (3,-1.2) {};
\node[right = 0.2cm of Cc] {Output Ports};

\end{tikzpicture} }
  \scalebox{0.9}{
  \begin{tikzpicture}[->, auto, semithick, draw=black, >=stealth,scale=1.5]
\tikzset{   
	outputport/.style = {circle, draw,thick,minimum size=1cm},
	outputportbis/.style = {circle, draw,thick},
}

\node[outputport] (O1) at (0,0) {O1};
\node[outputport] (O2) at (1,0) {O2}; 
\node[outputport] (O3) at (2,0) {O3}; 
\node[outputport] (O4) at (3,0) {O4}; 
\node[outputport] (O5) at (4,0) {O5}; 

\node[outputport] (O6) at (0,2) {O6};
\node[outputport] (O7) at (1,2) {O7}; 
\node[outputport] (O8) at (2,2) {O8}; 
\node[outputport] (O9) at (3,2) {O9}; 
\node[outputport] (O10) at (4,2) {O10}; 

\node[outputportbis,scale=0.9] (O11) at (1.8,0.7) {\footnotesize O12}; 
\node[outputportbis,scale=0.9] (O12) at (2.2,1.3) {\footnotesize O11}; 

\draw[red,ultra thick] (4.5,-0.1) -- (-0.8,-0.1) -- (-0.8,2.1) node[midway, left, red] {$f_r$} -- (3.25,2.1)  -- (3.4,2.5) ;
\node[circle,draw] (o1) at (3.4,2.5) {\scriptsize os1};
\draw[blue,dashed,ultra thick] (-0.5,2.2) -- (4.7,2.2) -- (4.7,-0.2) node[midway, right, blue] {$f_{bl}$} -- (0.45,-0.2) -- (0.3,-0.5);
\node[circle,draw] (o2) at (0.3,-0.5) {\scriptsize os2};
\draw[\dgreen,ultra thick,dash dot] (-0.5,1.8) -- (1,1.8) -- (1.8,0.55) node[midway, left, \dgreen] {$f_g$}  -- (1.8,0.1) -- (0.5,0.1) -- (0.4,0.5);
\node[circle,draw] (o3) at (0.4,0.5) {\scriptsize os3};
\draw[color={rgb:red,4;green,2;blue,1},ultra thick,dotted] (4.5,0.1) -- (3.1,0.1) -- (2.3,1.3) node[midway, right, color={rgb:red,4;green,2;blue,1}] {$f_{br}$}  -- (2.2,1.9) -- (3.5,1.9) -- (3.6,1.6);
\node[circle,draw] (o4) at (3.6,1.6) {\scriptsize os4};

\draw[black,ultra thick] (O6) -- (O7);
\draw[black,ultra thick] (O7) -- (O8);
\draw[black,ultra thick] (O8) -- (O9);
\draw[black,ultra thick] (O9) -- (O10);
\draw[black,ultra thick] (O10) -- (O5);
\draw[black,ultra thick] (O5) -- (O4);
\draw[black,ultra thick] (O4) -- (O3);
\draw[black,ultra thick] (O3) -- (O2);
\draw[black,ultra thick] (O2) -- (O1);
\draw[black,ultra thick] (O1) -- (O6);
\draw[black,ultra thick] (O5) -- (O4);
\draw[black,ultra thick] (O2) -- (o3);
\draw[black,ultra thick] (O2) -- (o2);
\draw[black,ultra thick] (O9) -- (o1);
\draw[black,ultra thick] (O9) -- (o4);
\draw[black,ultra thick] (O4) -- (O12);
\draw[black,ultra thick,scale=0.9] (O12) -- (O8);
\draw[black,ultra thick,scale=0.9] (O7) -- (O11);
\draw[black,ultra thick] (O11) -- (O3);
\end{tikzpicture}
}
  \caption{Top: a toy network with $4$ flows and with cyclic dependencies.
  Bottom: the graph of output ports induced by flow paths, also showing the flow paths.
  \label{fig:net}}
\end{figure}

The \emph{graph induced by flows} is the directed graph defined as follows:
\begin{enumerate}
\item Its vertices are output ports.
\item A directed edge exists from vertex $i$ to vertex $j$ if there is a communication link in the network from output port $i$ to the switch where $j$ resides and if there is at least one flow that uses this link and goes via $j$.
\end{enumerate}

We say that the network has cyclic dependencies if there is at least one cycle in the graph induced by flows.
In Figure~\ref{fig:net}, at the output port $O6$, three flows ($f_r$,$f_{bl}$, and $f_g$) compete in $O6$ to go to $O7$.
The delay-jitter at this node increases the burstiness of these flows.
This burstiness is then propagated to all next nodes on the considered flow path, 
hence the input burstiness of flow~$f_r$ at node $O6$ depends on the delay-jitter of node $O6$,
which also depends on the input burstiness of flow~$f_r$: this is a cyclic dependency. 

The toy example of Figure~\ref{fig:net} appears complex,  but it is chosen as a minimal example with two cycles,  as they are consistent with rules found in communication networks (for example a flow never loops). 
To this end, we need to introduce enough contention at output ports. 
If we suppress some switches in this network, for example the switch $O7$, then there is not enough contention: 
Contention will occur  at $O7$ only between flows $f_r$,$f_{bl}$, there will be direct links between $O6$ and $O8$ and $O6$ and $O12$,  and flow $f_g$ will not participate in a cyclic dependency with flows $f_r$,$f_{bl}$. 

If there is no cyclic dependency, the local stability condition mentioned in the previous section ensures that the network is stable,  and deterministic performance bounds can be computed at every note.  
In contrast, cyclic dependencies can make the network unstable in the sense that no finite bound exists for worst-case delay-jitter \cite{UnboundedAndrews09} even when local stability holds. Algorithms,  such as FPTFA,  and the algorithms in this paper compute delay-jitter bounds when they converge; when they diverge, it could be that the network is truly unstable or not.  
In practice, this latter case occurs when the network utilization is close to 100\%. 

\subsection{Notation List}
We use the following notation, illustrated on the example of Figure~\ref{fig:net}.
\begin{itemize}
\item $\calI$ is the set of nodes in the graph,  they correspond to output ports; $n=\abs{\calI}$.
\item Nodes are called \emph{terminal} when they lead to flow sinks (denoted with \textit{os} in Figure~\ref{fig:net}) 
\item Nodes are called \emph{non-terminal} when they lead to other nodes (denoted with \textit{O} in Figure~\ref{fig:net}).
\item Each flow $f$ has a path ($\paf(f)$) that is a sequence of connected nodes where the last element is terminal.  
We assume that a path has no loop, \ie, that a node appears at most one time in the path of~$f$.
For example, for flow $f_{br}$:
\begin{equation*}
\paf(f_{br}) = \left(O5,O4,O11,O8,O9,os4\right).
\end{equation*}
\item A flow $f$ is called \emph{fresh} at a node if this node is the first element of $\paf(f)$. 
For example, $f_{g}$ and $f_{bl}$ are fresh at node $O6$.
\item $\calL$ is the set of transit edges in the graph, \ie, the set of edges that lead to non-terminal nodes (transit links in Figure~\ref{fig:net}, top). 
In the physical network, it corresponds to the transit links, i.e. , the set of links that carry at least one transit flow (i.e., a flow that is neither fresh nor terminal).
For example,
\begin{center}
$(O8,O9)\in\calL$ but $(O9,os1)\nin\calL$.
\end{center}
\item $\forall i \in \calI$, $\inc(i) \subset \calL$ is the set of transit edges that are incident to node $i$.  
For example, 
\begin{equation*}
\inc(O6)=\lc (O1,O6)\rc. 
\end{equation*}
\item $\forall i \in \calI$, $\out(i) \subset \calL$ is the set of transit edges that leave node $i$.  
For example, 
\begin{equation*}
\out(O7)=\lc(O7,O8),(O7,O12)\rc. 
\end{equation*}
\item $\forall \ell \in \calL$, $\pred(\ell)$ denotes the set of nodes that are crossed by at least one flow that is present in $\ell$, and that are upstream of $\ell$ for such flows.
In other words, node $i\in\pred(\ell)$ if and only if there exists one flow~$f$ such that node~$i$ and edge $\ell$ are on the path of flow~$f$, in this order.
For example, for $\ell=(O3,O2)$, 
\begin{equation*}
\pred(\ell)  = \lc O3, O4, O5,O6,O7,O8,O9,O10,O12\rc.
\end{equation*}
We also introduce $\pred_f(\ell)$ as the set of nodes that are crossed by flow~$f$ and that are upstream of $\ell$.

For $\ell=(O3,O2)$, and flow $f_r$:
\begin{equation*}
\pred_{f_r}(\ell)  = \lc O3, O4, O5 \rc.
\end{equation*}
%
\item Recall that flow $f$ is constrained at the source by a leaky bucket with rate $r_f$ and burstiness $b_f$.

\end{itemize}

\par\noindent The algorithms in this paper estimate bounds on delay-jitter and propagated burstiness.  
The notation for these is as follows.
\begin{itemize}
\item $\forall \ell$ $\in$ $\calL$, $z_{\ell}$ is a vector that has one component for every transit flow that is carried by $\ell$.
Every component is an upper bound on the propagated burstiness of the flow on the transit link $\ell$ (denoted with $b_{\ell}^{f}$).

With $\ell$=$(O6,O7)$:
\begin{equation*}
z_\ell = \left(b_{(O6,O7)}^{f_r},b_{(O6,O7)}^{f_g},b_{(O6,O7)}^{f_{bl}}\right). 
\end{equation*}
\item $\forall M \subset \calL$, $z_{M}$ is the set of $z_{\ell}$ for all $\ell \in M$.  
In particular, $z_{\calL}$ denotes the collection of all bounds on propagated burstiness.
\item $\forall i \in \calI$, $d_i$ is a vector that is meant to contain an upper bound on the delay-jitters at node $i$ for every flow $f$ (transit or not) that uses node $i$.
The delay-jitter at a node is defined as the worst-case delay minus the best-case delay;
it is computed per bit, from entrance to the packetizer to exit out of the output port.
The delay includes the packetization delay; if all flows have maximum and minimum packet length and all links have same rate,
then all flows have the same delay-jitter bound and $d_i$ is a single number;
else the delay will depend on the input port and the packet sizes,
hence depends on the flow, not just the output port $i$ \cite{mohammadpour2019improved}. 
\item $\forall J\subset \calI$, $d_J$ denotes the collection of all $d_i$ for $i \in J$.  
In particular, $d_{\calI}$ is the collection of all delay-jitter bounds.

\end{itemize}
 
The following operators are introduced in subsequent sections and are given here for completeness.

~\\
\par\noindent
\begin{tabular}{|c|c|}

\hline

$\calD$ & \begin{minipage}[t]{8cm}
$d=\calD(z)$ is a vector of delay-jitter bounds derived from propagated burstinesses 
(Section~\ref{sec:tfa}, used by all algorithms).
\end{minipage}
\\

\hline

$\calG$ & \begin{minipage}[t]{8cm}
$\calG(z,d) \eqdef (\calY(d),\calD(z))$ (Section~\ref{sec:sync-tfa-def}).
\end{minipage} 

\\

\hline
$\calY$ & \begin{minipage}[t]{8cm}
$z=\calY(d)$ is a vector of propagated burstinesses derived from delay-jitter bounds (Section~\ref{sec:sync-tfa-def}, used by \synctfa, \asynctfa and \asyncalttfa).
\end{minipage} 

\\

\hline
$\calZ$ & \begin{minipage}[t]{8cm}
$z'=\calZ(d,z)$ is a vector of output burstinesses derived from delay-jitter bounds and from input burstinesses (Section~\ref{sec:tfa}, used by TFA and FPTFA).
\end{minipage} 

\\

\hline
\end{tabular}

~\\
Beyond delay-jitter, the deterministic performance bounds of delay and backlog are derived from $z_{\calL}$ and from fixed parameters of the network, hence we do not need to consider them further. 

\section{State-of-the-Art Algorithms: TFA~\cite{tfa_disco} and \fptfa~\cite{thomas2019cyclic}}
\label{sec:SoA}

In this section, we present two state-of-the-art algorithms that compute worst case bounds: 
The former, Total Flow Analysis (TFA)~\cite{tfa_disco}, is used for networks without cyclic dependencies on the graph induced by flow path; 
the latter,  Fixed-Point Total Flow Analysis (\fptfa)~\cite{thomas2019cyclic}, applies to cases with cyclic dependencies.  
Our presentation is more compact than in the original references and is used to support the theoretical analysis in this paper.

\subsection{Total Flow Analysis (TFA)~\cite{tfa_disco}} 
\label{sec:tfa}

\begin{figure}
  \centering
	\begin{tikzpicture}[->, auto, semithick, draw=black, >=stealth,scale=1.75]
    \scalebox{0.85}{  \input{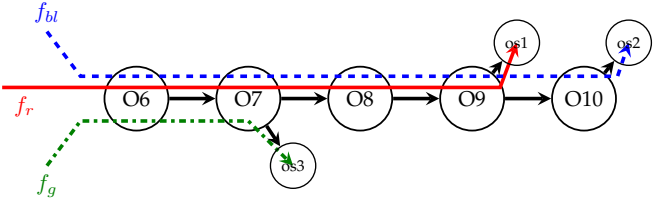} 
    }
\end{tikzpicture}
     \caption{A Feed-forward network with three flows $f_{r}$,$f_{bl}$ and $f_{gr}$ on which TFA is applied. \label{fig:feedforward}}
\end{figure}

The Total Flow analysis (TFA) algorithm~\cite{tfa_disco} performs worst-case analysis on graphs without cyclic dependencies (see an example of feed-forward network on Figure~\ref{fig:feedforward}).
TFA uses network calculus to analyse each node one after the other, starting with nodes that have only fresh input flows (such nodes must exist when there is no cyclic dependency).  
For each node,  TFA computes the delay-jitter bound and the output-burstinesses bound. 
The output burstinesses are then used as input for the following nodes,  in the sense of the flows.  
Iteratively TFA computes all delay-jitter bounds and burstinesses bounds,  for all nodes and links of the network.

TFA~\cite{tfa_disco} bounds are then improved by~\cite{mifdaoui2017beyond} with TFA++ by taking into account the effect of line-shaping constraints at the input and the output of each node. 
Then \cite{thomas2019cyclic} obtains a tighter delay-jitter bound within nodes by considering the effect of packetizer. 

The first step of the algorithm is to create a \textit{topological order}~\cite{Cormen_ToplogicalOrder} of the acyclic directed graph by labelling the nodes of the network $i_1, ..., i_n$.  
It means that the input of a node $i_k$ belongs to the output of the previous nodes $\{i_1, \cdots i_{k-1}\}$. Specifically, $\inc(i_1) = \emptyset$ (all flow that enters in node $i_1$ are fresh) and 
\begin{equation}
\forall k \in \{2,\cdot,n\}, \, \inc(i_k) \subset \out(i_1) \cup \cdots \cup \out(i_{k-1}). \label{set:iktfa}
\end{equation}
Let us take the example of the feed-forward network in Figure~\ref{fig:feedforward}. 
This network is composed of three flows $f_{r}$,$f_{bl}$,$f_{gr}$, that first cross node $O6$.
The only possible labelling is $\{O6,O7,O8,O9,O10\}$.

Then the second step is the Algorithm~\ref{alg:TFA} that performs an analysis of each node in the order of the labelling by computing delay-jitter and output bursts. 

\begin{algorithm}[H]
	\caption{Total Flow Analysis (TFA)}
	\begin{algorithmic}[1]
		\State  $z_{\ell} \gets 0$, $\forall\ell \in \calL$;
		\State $d_i \gets 0$, $\forall i\in \calI$;
		\For{$k \leftarrow 1 $ to $n$}
			\State $d_{i_k}\gets \calD_{i_k}\lp z_{\inc(i_k)} \rp$;
			\State $\forall \ell \in \out(i_k)$, $z_\ell =\calZ_{i_k} \lp d_{i_k}, z_{\inc(i_k)} \rp$;
		\EndFor
		\State $print(z,d)$;
	\end{algorithmic}
	\label{alg:TFA}
\end{algorithm}

To perform this analysis, TFA uses network calculus methods to derive delay-jitter and burstiness bounds at every node, abstracted as follows:
\begin{itemize}
\item $\forall i \in \calI$, $\calD_i$ is the function that gives a delay-jitter bound $d_i$ for a node~$i$.  
It is defined as follows: $d_i \eqdef \calD_i \lp z_{\inc(i)} \rp$, where $z_{\inc(i)}$ is the burstiness bounds of every transit flow that enters into $i$. 
A formulation of the $\calD_i$ function is given in~\cite[Eq.(3)]{thomas2019cyclic}. 
Note that $\calD_i$ accounts for the burstiness of all flows, transit or fresh, but only the burstiness of transit flows is captured by the argument,
as the burstiness at the source is assumed to be fixed.
As shown in~\cite[Eq.(3)]{thomas2019cyclic}, the delay-jitter at each node is the minimum of affine functions in the input burstinesses,
hence $\forall i \in \calI$, every coordinate of $\calD_i$ is concave.

For example,  in Figure~\ref{fig:feedforward}, the delay-jitter at node $i=O6$ is 
\begin{equation}
d_{O6} = d_{O6}^{f_r} =d_{O6}^{f_g} =d_{O6}^{f_{bl}} = \calD(z_{\texttt{inputO6}})
\end{equation}
with $z_{\texttt{inputO6}}$ the input vector of burstinesses at node $O6$ defined in Eq.~\eqref{eq:inBurst}.
\begin{equation}
z_{\texttt{inputO6}}=\left(b^{f_r}_{\texttt{inputO6}},b_{\texttt{inputO6}}^{f_{bl}},b_{\texttt{inputO6}}^{f_{g}} \right).
\label{eq:inBurst}
\end{equation} 
\item $\forall i\in \calI$, $\calZ_{i}$ is the mapping that provides a collection of burstiness bounds for all transit flows at the output of node $i$.
It is defined as follows:  $z'_{\out(i)} \eqdef \calZ_i\lp d_i, z_{\inc(i)} \rp$ where $d_i$ is the delay-jitter bound of node~$i$ and $z_{\inc(i)}$ is the input-burstiness bounds at this node.
Every coordinate of $\calZ_{i}$ is affine in its arguments,  as propagated burstiness is simply equal to the input burstiness plus the rate of the flow times the delay-jitter for the flow at this node. 

For example in Figure~\ref{fig:feedforward},  the output burstiness vector at node $O6$ is:
\begin{eqnarray*}
z'_{(O6,O7)} &=& \calZ_i\lp d_{O6}, z_{\inc(O6)} \rp \\
&=& (b^{f_r}_{(O6,O7)},b^{f_{bl}}_{(O6,O7)},b^{f_{gr}}_{(O6,O7)}),
\end{eqnarray*} 
Note that in this example $\inc(O6)=\emptyset$,  $z'_{(O6,O7)}$ depends on the burstinesses at source :
\begin{eqnarray}
&&b^{f_{r}}_{(O6,O7)} = b^{f_{r}}_{\texttt{inputO6}} + r_{f_{r}} d_{O6}\\
&&b^{f_{bl}}_{(O6,O7)} = b^{f_{bl}}_{\texttt{inputO6}} + r_{f_{bl}} d_{O6}\\
&&b^{f_{gr}}_{(O6,O7)} = b^{f_{gr}}_{\texttt{inputO6}} + r_{f_{gr}} d_{O6}
\end{eqnarray}
\end{itemize}
At the output of node $O6$,  with function $\calD$ and $\calZ$,  delay-jitter and output burstiness bounds have been computed (Lines~4-5 of Algorithm~\ref{alg:TFA}).
Then the same analysis is performed for all the other nodes in sequence $\{O7,O8,O9,O10\}$ to finish the worst-case analysis.

Let $\calZ$ and $\calD$ be compact notations for these mappings, \ie, 
\begin{eqnarray}
\lp z'= \calZ(d,z) \rp & \Leftrightarrow & \lp z'_{\out(i)}=\calZ_{i}\lp d_{i},z_{\inc(i)} \rp, \forall i \in \calI\rp\\
\lp d= \calD(z) \rp & \Leftrightarrow &\lp d_i=\calD_i\lp z_{\inc(i)}\rp, \forall i\in \calI\rp
\end{eqnarray}
The local stability conditions ensures that  $\calD$ is well defined, \ie,  returns finite values for all finite values of its arguments.

\subsection{Fixed-Point Total Flow Analysis (FPTFA)~\cite{thomas2019cyclic}}
\label{sec:fptfa}

In the case where we have cyclic dependencies,  such as in Figure~\ref{fig:net},  \cite{thomas2019cyclic} introduces another algorithm: \fptfa. 
Here we describe this algorithm in a different way of~\cite{thomas2019cyclic}, in a form that is more compact and suitable for analysis. 

\fptfa uses a cut $L \subset \calL$ such that the resulting graph has no cycle when the edges in $L$ are removed, and it iterates on $z_L$, which is the vector of estimates of the burstinesses at the cut. 
For example, in Figure~\ref{fig:net},  there are several cyclic dependencies, and $L$ must contain at least two links. 
One possible cut is 
\begin{equation}
L = \{(O1,O6),(O10,O5)\}
\end{equation} 
and then:
\begin{equation}
  z_{L}=\left( z_{(O1,O6)}^{br},z_{(O10,O5)}^{bl} \right) .
\end{equation} 
Other cuts are possible,  for example,  $ L=\{(O1,O6),(O11,O8)\}$.  
These two minimal cuts are represented with red arrows in Figure~\ref{fig:cut}. Note that FPTFA does not require the cuts to be minimal, and works with any cut that leaves the network free of cyclic dependencies. Indeed, as argued in \cite{thomas2019cyclic}, finding minimal cuts may be the most time-consuming part of the method; in some topologies (such as rings of rings), it is easy to find valid cuts (e.g. by cutting every ring at one link). Such cuts are not guaranteed to be minimal, as the minimal cuts depend not just on the topology but also the flow paths. 
As an extreme, it is possible to take $L=\calL$, i.e. , to cut all links.

\def\opac{0.6}
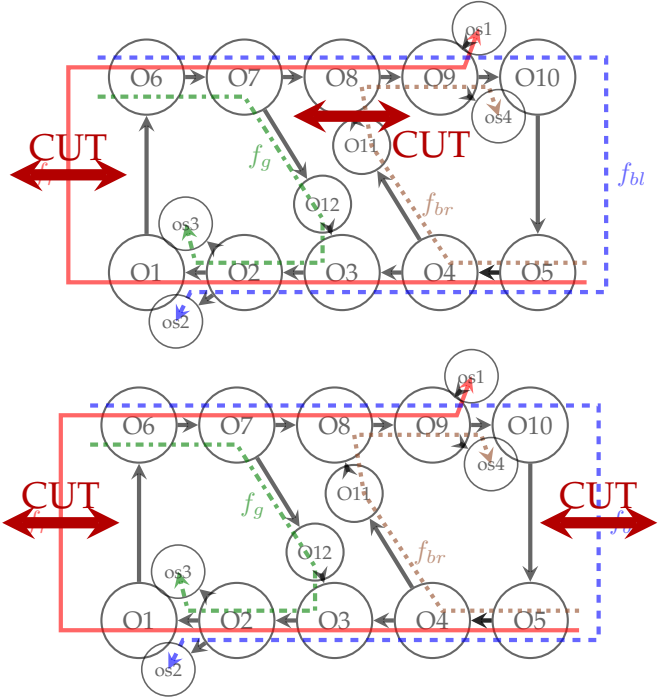
\begin{figure}
  \centering
   \begin{tikzpicture}[->, auto, semithick, draw=black, >=stealth,scale=1.3] 
   \begin{scope}[opacity=\opac,text opacity=\opac]
  \tikzset{   
	outputport/.style = {circle, draw,thick,minimum size=1cm},
	outputportbis/.style = {circle, draw,thick},
}

\node[outputport] (O1) at (0,0) {O1};
\node[outputport] (O2) at (1,0) {O2}; 
\node[outputport] (O3) at (2,0) {O3}; 
\node[outputport] (O4) at (3,0) {O4}; 
\node[outputport] (O5) at (4,0) {O5}; 

\node[outputport] (O6) at (0,2) {O6};
\node[outputport] (O7) at (1,2) {O7}; 
\node[outputport] (O8) at (2,2) {O8}; 
\node[outputport] (O9) at (3,2) {O9}; 
\node[outputport] (O10) at (4,2) {O10}; 

\node[outputportbis,scale=0.9] (O11) at (1.8,0.7) {\footnotesize O12}; 
\node[outputportbis,scale=0.9] (O12) at (2.2,1.3) {\footnotesize O11}; 

\draw[red,ultra thick] (4.5,-0.1) -- (-0.8,-0.1) -- (-0.8,2.1) node[midway, left, red] {$f_r$} -- (3.25,2.1)  -- (3.4,2.5) ;
\node[circle,draw] (o1) at (3.4,2.5) {\scriptsize os1};
\draw[blue,dashed,ultra thick] (-0.5,2.2) -- (4.7,2.2) -- (4.7,-0.2) node[midway, right, blue] {$f_{bl}$} -- (0.45,-0.2) -- (0.3,-0.5);
\node[circle,draw] (o2) at (0.3,-0.5) {\scriptsize os2};
\draw[\dgreen,ultra thick,dash dot] (-0.5,1.8) -- (1,1.8) -- (1.8,0.55) node[midway, left, \dgreen] {$f_g$}  -- (1.8,0.1) -- (0.5,0.1) -- (0.4,0.5);
\node[circle,draw] (o3) at (0.4,0.5) {\scriptsize os3};
\draw[color={rgb:red,4;green,2;blue,1},ultra thick,dotted] (4.5,0.1) -- (3.1,0.1) -- (2.3,1.3) node[midway, right, color={rgb:red,4;green,2;blue,1}] {$f_{br}$}  -- (2.2,1.9) -- (3.5,1.9) -- (3.6,1.6);
\node[circle,draw] (o4) at (3.6,1.6) {\scriptsize os4};

\draw[black,ultra thick] (O6) -- (O7);
\draw[black,ultra thick] (O7) -- (O8);
\draw[black,ultra thick] (O8) -- (O9);
\draw[black,ultra thick] (O9) -- (O10);
\draw[black,ultra thick] (O10) -- (O5);
\draw[black,ultra thick] (O5) -- (O4);
\draw[black,ultra thick] (O4) -- (O3);
\draw[black,ultra thick] (O3) -- (O2);
\draw[black,ultra thick] (O2) -- (O1);
\draw[black,ultra thick] (O1) -- (O6);
\draw[black,ultra thick] (O5) -- (O4);
\draw[black,ultra thick] (O2) -- (o3);
\draw[black,ultra thick] (O2) -- (o2);
\draw[black,ultra thick] (O9) -- (o1);
\draw[black,ultra thick] (O9) -- (o4);
\draw[black,ultra thick] (O4) -- (O12);
\draw[black,ultra thick,scale=0.9] (O12) -- (O8);
\draw[black,ultra thick,scale=0.9] (O7) -- (O11);
\draw[black,ultra thick] (O11) -- (O3);
   \end{scope}
   \draw[red!70!black,line width=1.5mm, <->]  (-1.4,1)-- (-0.2,1) node[red!70!black,above,midway] {\Large CUT};
   \draw[red!70!black,line width=1.5mm, <->]  (1.5,1.6) --  (2.7,1.6) node[red!70!black,below] {\,\,\,\,\,\,\,\,\,\,\Large CUT};
   \end{tikzpicture}
   \begin{tikzpicture}[->, auto, semithick, draw=black, >=stealth,scale=1.3] 
   \begin{scope}[opacity=\opac,text opacity=\opac]
	\tikzset{   
	outputport/.style = {circle, draw,thick,minimum size=1cm},
	outputportbis/.style = {circle, draw,thick},
}

\node[outputport] (O1) at (0,0) {O1};
\node[outputport] (O2) at (1,0) {O2}; 
\node[outputport] (O3) at (2,0) {O3}; 
\node[outputport] (O4) at (3,0) {O4}; 
\node[outputport] (O5) at (4,0) {O5}; 

\node[outputport] (O6) at (0,2) {O6};
\node[outputport] (O7) at (1,2) {O7}; 
\node[outputport] (O8) at (2,2) {O8}; 
\node[outputport] (O9) at (3,2) {O9}; 
\node[outputport] (O10) at (4,2) {O10}; 

\node[outputportbis,scale=0.9] (O11) at (1.8,0.7) {\footnotesize O12}; 
\node[outputportbis,scale=0.9] (O12) at (2.2,1.3) {\footnotesize O11}; 

\draw[red,ultra thick] (4.5,-0.1) -- (-0.8,-0.1) -- (-0.8,2.1) node[midway, left, red] {$f_r$} -- (3.25,2.1)  -- (3.4,2.5) ;
\node[circle,draw] (o1) at (3.4,2.5) {\scriptsize os1};
\draw[blue,dashed,ultra thick] (-0.5,2.2) -- (4.7,2.2) -- (4.7,-0.2) node[midway, right, blue] {$f_{bl}$} -- (0.45,-0.2) -- (0.3,-0.5);
\node[circle,draw] (o2) at (0.3,-0.5) {\scriptsize os2};
\draw[\dgreen,ultra thick,dash dot] (-0.5,1.8) -- (1,1.8) -- (1.8,0.55) node[midway, left, \dgreen] {$f_g$}  -- (1.8,0.1) -- (0.5,0.1) -- (0.4,0.5);
\node[circle,draw] (o3) at (0.4,0.5) {\scriptsize os3};
\draw[color={rgb:red,4;green,2;blue,1},ultra thick,dotted] (4.5,0.1) -- (3.1,0.1) -- (2.3,1.3) node[midway, right, color={rgb:red,4;green,2;blue,1}] {$f_{br}$}  -- (2.2,1.9) -- (3.5,1.9) -- (3.6,1.6);
\node[circle,draw] (o4) at (3.6,1.6) {\scriptsize os4};

\draw[black,ultra thick] (O6) -- (O7);
\draw[black,ultra thick] (O7) -- (O8);
\draw[black,ultra thick] (O8) -- (O9);
\draw[black,ultra thick] (O9) -- (O10);
\draw[black,ultra thick] (O10) -- (O5);
\draw[black,ultra thick] (O5) -- (O4);
\draw[black,ultra thick] (O4) -- (O3);
\draw[black,ultra thick] (O3) -- (O2);
\draw[black,ultra thick] (O2) -- (O1);
\draw[black,ultra thick] (O1) -- (O6);
\draw[black,ultra thick] (O5) -- (O4);
\draw[black,ultra thick] (O2) -- (o3);
\draw[black,ultra thick] (O2) -- (o2);
\draw[black,ultra thick] (O9) -- (o1);
\draw[black,ultra thick] (O9) -- (o4);
\draw[black,ultra thick] (O4) -- (O12);
\draw[black,ultra thick,scale=0.9] (O12) -- (O8);
\draw[black,ultra thick,scale=0.9] (O7) -- (O11);
\draw[black,ultra thick] (O11) -- (O3);
   \end{scope}
   \draw[red!70!black,line width=1.5mm, <->]  (-1.4,1)-- (-0.2,1) node[red!70!black,above,midway] {\Large CUT};
    \draw[red!70!black,line width=1.5mm, <->]  (4.1,1)-- (5.3,1) node[red!70!black,above,midway] {\Large CUT};
   \end{tikzpicture}
     \caption{Different versions of \fptfa with two different cuts. 
     Top figure: the cut is $ L=\{(O1,O6),(O11,O8)\}$;
     the flows $f_r$ and $f_{br}$ are cut. 
     Bottom figure: the cut is $  L = \{(O1,O6),(O10,O5)\} $; the flows $f_r$ and $f_{bl}$ are cut. 
  \label{fig:cut}}
\end{figure}

The graph obtained when removing the edges in $L$ has no cycle,  hence it is possible to label the nodes correspondingly in a feed-forward manner as done in \tfa.
For a given cut $L$, we thus build a \textit{topological order}~\cite{Cormen_ToplogicalOrder},  and label nodes with indexes $i_1, ..., i_n$ accordingly to this order.  
This labelling is defined by Eq.~\eqref{set:i1}-\eqref{set:ik}.
\begin{subequations}
\begin{empheq}{align}
&\inc(i_1) \subset L \label{set:i1} \\
&\forall k \in \{2,\cdot,n\}, \, \inc(i_k) \subset L\cup \out(i_1) \cup \cdot \cup \out(i_{k-1}) \label{set:ik}
\end{empheq}
\end{subequations}
In the example of the bottom figure in Figure~\ref{fig:cut},  a possible node labelling is
\begin{equation*}
(O6,O7,O5,O4,O11,O12,O8,O3,O9,O2,O10,O1).
\end{equation*}
Note that, in this paper, $\inc()$ and $\out()$ always refer to the original, uncut, graph. 

\fptfa is described in Algorithm~\ref{alg:FPTFA}. 
In one iteration, \fptfa computes a new value $z'_L$ of the vector of burstinesses at the cut and then  
prints the value of the delay-jitter and burstiness bounds $(z,d)$ valid for the cut network.
We call $\calF^L$ the mapping that transforms the vector of burstinesses $z_L$ into $z'_L$, specifically, as
described in lines 4-8 of Algorithm~\ref{alg:FPTFA}. This is the same as
the function called $\calF\calF$ in~\cite{thomas2019cyclic}, where we highlight the dependency on the cut~$L$.
It follows that \fptfa computes the successive iterated of $\calF^L$, starting with initial value $z_L=0$.

\begin{algorithm}[htb]
	\caption{Fixed-Point TFA (\fptfa)}
	\begin{algorithmic}[1]
		\State  $z_{\ell} \gets 0$, $\forall\ell \in \calL$;
		\State $d_i \gets 0$, $\forall i\in \calI$;
		\While{\texttt{true}}
		\For{$k \leftarrow 1 $ to $n$}
		\State $d_{i_k}\gets \calD_{i_k}\lp z_{\inc(i_k)} \rp$;
		\State $\forall \ell \in \out(i_k) \setminus L$, $z_\ell =\calZ_{i_k} \lp d_{i_k}, z_{\inc(i_k)} \rp$;
		\State $\forall \ell \in \out(i_k) \cap L$, $z'_\ell =\calZ_{i_k} \lp d_{i_k}, z_{\inc(i_k)} \rp$;
		\EndFor
		\State $z_{\ell} \gets z'_{\ell}$, $\forall\ell \in L$ ;
		\State $print(z,d)$;
		\EndWhile
	\end{algorithmic}
	\label{alg:FPTFA}
\end{algorithm}

Theorem 2 in~\cite{thomas2019cyclic} proves that if the successive iterated of the burstinesses at cut $z_L$ converge, and if the network is empty at time $0$,
then values of $d$ and $z$ computed by \fptfa are valid bounds for the original, non cut, network.

We notice that when the network is without cyclic dependencies and the cut of \fptfa is $L=\emptyset$,  then \fptfa and \tfa compute the  same bounds.  
Indeed, indefinitely \fptfa prints  at each iteration the same burstinesses and delays bounds as \tfa outputs.

Back to a network with cyclic dependency, a question that arises is whether 
the choice of a cut influences the convergence and the value of the bounds computed by \fptfa. 
We answer this question in the following section by introducing a new algorithm \synctfa.  
We prove that  \fptfa with cut $L$ and \synctfa compute the same bounds, therefore that the choice of cut has no influence on the end-result.

\section{Synchronous TFA}
\label{sec:sync-tfa}

We now introduce \synctfa, a new algorithm that simultaneously updates delay-jitter bounds at every node and propagated burstinesses at all transit links. 
It applies to networks without or with cyclic dependencies. 
In the former case, it stops after a number of iterations and is equivalent to TFA.
In the latter case, it iterates until it either finds a fixpoint or it diverges.  
The particularity of this new algorithm is that, contrary to \fptfa, it does not require cutting the network.

\subsection{Definition of \synctfa }
\label{sec:sync-tfa-def}
\synctfa updates the delay-jitter bounds by using the same method as TFA and \fptfa but for, propagated burstiness, uses a method that slightly differs, and that we now describe. 
We need to introduce $\forall \ell\in \calL$, $\calY_{\ell}$ a mapping such that 
$z_{\ell} \eqdef \calY_{\ell}\lp d_{\pred(\ell)} \rp$ is a collection of valid burstiness bounds
for all transit flows present in edge $\ell$, when the delay-jitters at all upstream nodes are given by $d_{\pred(\ell)}$.
Every coordinate of $\calY_{\ell}$ is a propagated burstiness; it is equal to the burstiness at the source (considered to be a constant) plus the rate of the flow times the accumulated delay-jitter bound on the path of the flow.  
For example, for $\ell=(O3,O2)$, the first coordinate of $\calY_{\ell}$ returns
\begin{equation*}
z_{\ell}^1=b_{(O3,O2)}^{f_g}=r_{f_g}\sum_{i\in\lc O3,O12,O7,O6\rc}d_i^{f_g} + b_{S2}^{f_g}.
\end{equation*}
In contrast, $\calZ$, used in \tfa and \fptfa, estimates the output burstiness at a node from the delay-jitter bounds at this node and the input burstiness.

Note that $\calY_{\ell}$ is affine.

Also introduce $\calG$, such that $\calG(z,d) \eqdef (\calY(d),\calD(z))$. 
This function is isotonic and concave by definition of function $\calD$ in~\cite[Eq.(3)]{thomas2019cyclic} and because $\calY$ is affine.  
We can now define \synctfa as follows:

\begin{algorithm}[H]
	\caption{Synchronous TFA (\synctfa)}
	\begin{algorithmic}[1]
		\State $z_\ell \gets 0$, $\forall\ell \in \calL$; 
		\State $d_i \gets 0$, $\forall i\in \calI$;
		\While{\texttt{true}}
		\State $(z,d)\gets (\calG(z,d))$;
		\State $print(z,d)$;
		\EndWhile
	\end{algorithmic}
		\label{alg:SyncTFA}
\end{algorithm}

The iteration $k$ of \synctfa prints $(z^k,d^k)=\calG^k (0, 0)$.

\subsection{Correctness of \synctfa }
In this Section we show the correctness of \synctfa, which means that if the successive iterates of \synctfa converge, they provide valid bounds for 
propagated burstiness and delay-jitter. This is a similar result as Theorem 2 in~\cite{thomas2019cyclic} (which is for FPTFA), but our method of proof is simpler and does not require any assumptions on propagation delays or initial network state.

We use the time-stopping method~\cite{LebBook}, which can be cast as follows. Consider that the network starts at
time $0$ in some arbitrary state and fix some time $\tau\geq 0$. Modify the sources such that they stop sending after
time $\tau$ and call $\calN^\tau$ the resulting modified network. Since all sources are constrained at the sources and $\tau$ is finite, the number of bits that ever exist in $\calN^\tau$ is finite. Therefore, the worst-case delays and propagated burstinesses, $d^{\tau}$ and $z^{\tau}$ are finite\footnote{The worst case burstiness for flow $f$ at some point in the network is defined as $\sup_{0\leq s, t} (R(t)-R(s)-r_f(t-s))$ where $R(t)$ is the number of bits observed at this point for flow $f$ between times $0$ and $t$.}. By network calculus, $\calY_\ell(d)$ provides valid burstiness bounds at every transit link $\ell$ and $\calD_i(z)$ provides valid burstiness bounds at every node $i$ whenever $d$ and $z$ are also valid bounds. Therefore, in $\calN^\tau$, $\calY_\ell(d^\tau)$ and $\calD_i(z^\tau)$ are valid bounds. Since $d_i^\tau$ and $z_\ell^\tau$ are minimal bounds, it follows that $z_\ell^\tau\leq \calY_\ell(d^\tau)$ and $d_i^\tau\leq \calD_i(z^\tau)$ for every node $i$ and transit link $\ell$. In compact form:
\begin{equation}
	(z^\tau, d^\tau)\leq \calG(z^\tau, d^\tau)
\end{equation}
Let us introduce the following set:
$$\lo(\calG) \eqdef \lc (z,d)\mst (z,d)\leq \calG(z,d)\rc.$$
It follows that, for every $\tau\geq 0$
\begin{equation}
	(z^\tau, d^\tau)\in \lo(\calG)
	\label{eq-ts1}
\end{equation}

Back to the original network, let $(z(t), d(t))$ be the worst-case burstinesses and delays observed in the interval $[0,t]$, where $t\geq 0$.
By causality, in any network, the worst-case delays and burstinesses that can be observed in the time interval $[0, t]$
depend only on the history of the network up to time $t$. It follows that
\begin{equation}
	(d(t),z(t))\leq (d^\tau, z^\tau) \mbox{ if } t\leq \tau
	\label{eq-ts2}
\end{equation}
Combining \eqref{eq-ts1} and \eqref{eq-ts2} we see that, if $\lo(\calG)$ is bounded, an upper-bound on  $\lo(\calG)$ provides valid bounds for the original network. This explains why sets such as $\lo(\calG)$ play an important role in the analysis of the algorithms in this paper.

The following theorem establishes a proof of correctness of \synctfa. 

\begin{theorem}
If $\lo(\calG)$ is bounded,  then 
\begin{enumerate}
\item $\calG$ has a unique fixpoint $(\bar{z},\bar{d})$.
\item $\bar{d}$ is a set of valid delay-jitter bounds at all nodes and $\bar{z}$ gives valid burstiness bounds at all transit links.
\item The \synctfa sequence $(z^k,d^k)$ converges to $(\bar{z},\bar{d})$.
\end{enumerate}
%
If $\lo(\calG)$ is unbounded, then $\calG$ has no fixpoint and the \synctfa sequence does not converge.
\label{th:lowGbounded}
\end{theorem}
It follows from Theorem~\ref{th:lowGbounded} that if the \synctfa sequence $(z^k,d^k)$ converges, the limit gives a set of valid delay-jitter bounds at all nodes.
\begin{proof}
$\calG$ is isotonic, \ie,  $\calG(z,d)\leq\calG (z',d')$ whenever $z\leq z'$ and $d\leq d'$ (comparison is coordinatewise).  
Now $(0,0)=(z^0,d^0)\leq (z^1,d^1)$,  hence $(z^k,d^k)\leq (z^{k+1},d^{k+1})$ for all $k\geq 0$.  
Furthermore,  $(z^k,d^k)\in \lo(\calG)$ because $(z^{k+1},d^{k+1})=\calG(z^k,d^k)$.  
Also, $\calG$ is concave because $\calY$ is affine and $\calD$ is also affine by the definition of $\calD$ in~\cite[Eq.(3)]{thomas2019cyclic}. 
Last, $\calG(0) > 0$,  because 
\begin{itemize}
\item There is a packetizer at each input port and $\ell_{max} > 0$,  thus $\calD(0)>0$.
\item By Theorem~1 in~\cite{thomas2019cyclic} the burstiness increase after a packetizer is $\frac{r}{c}\ell_{max} >0$ (with $r$ the rate of the flow aggregate and $c$ the transmission rate of the line at the input port), thus $\calY(0) > 0$.
\end{itemize}

Now assume that $\lo(\calG)$ is bounded. 
By Corollary~\ref{coro-c1}-1), $\calG$ has a unique fixpoint, say $(\bar{z},\bar{d})$, which proves 1).
Also $(z(t),d(t)) \leq (z^t,d^t)\in\lo(\calG)$ by the time-stopping method and,  
by Corollary~\ref{coro-c1}-1, $(\bar{z},\bar{d})$ is the largest element of $\lo(\calG)$; this proves 2). 
Also, $(z^k,d^k)\leq (\bar{z},\bar{d})$, hence the sequence converges to a finite limit. 
Since $\calG$ is continuous, the limit is a fixpoint of $\calG$; 
by uniqueness of the fixpoint, the limit is $(\bar{z},\bar{d})$, which proves 3).

If $\lo(\calG)$ is unbounded,  then by Corollary~\ref{coro-c1}-2), $\calG$ has no fixpoint and diverges.
\end{proof}

\synctfa , like all algorithms in this paper,  is built for networks with cyclic dependencies, but also applies to networks without cyclic dependencies.  
When the network has no cyclic dependencies 
Theorem~\ref{th:level} establishes that \synctfa is stationary after a finite number of iteration and that it provides valid delay-jitter bounds.

To define Theorem~\ref{th:level}, we use the notion of level number~\cite{level}, defined as follows. 
The \emph{level number} of a node~$i$, denoted with $\dep(i)$, 
is the length of the longest directed path to this node on the graph induced by flows. Here, note that a path is a concatenation of adjacent edges in the graph induced by flows. It need not be a path followed by a flow of the original network.

The following properties hold for the level numbers:
\begin{enumerate}
\item If a node $i \in \calI$ has a level number of~$k=1$,  then there are only fresh flows at $i$.  
Thus for this node~$i$,   we have $\inc(i) = \emptyset$.
\item If node $i \in \calI$ has level number $k$,  then: 
\begin{eqnarray}
\forall f \text{ s.t. } i \in \paf(f),  \forall j \in \pred_f((\cdot,i)),  \hspace{2cm} \nonumber\\
\hspace{3cm} \dep(j) \leq k-1
\label{eq:defprof}
\end{eqnarray}
\end{enumerate}  
The \textit{number of levels} of a network is the maximum level number of all its nodes.  

For example, in the bottom of Figure~\ref{fig:cut},  the graph induced by flows has 6 levels:  
Nodes with level number $1$ are $O6$ and $O5$.  
$O1$ and $O10$ have level number~$6$.
They have a respective longest directed path $(O6,O7,O12,O3,O2,O1)$ and $(O5,O4,O11,O8,O9,O10)$.
\begin{theorem}
In a network without cyclic dependencies and with $k$ levels, \synctfa is stationary in at most $2k$ steps.
\label{th:level}
\end{theorem}
It follows from Theorems \ref{th:lowGbounded} and~\ref{th:level}, that, in a network without cyclic dependencies, $\lo(\calG)$ is bounded and \synctfa always computes valid delay-jitter and burstiness bounds.
\begin{proof}
We note $z^{(j)}$ the value of $z$ after $j$ iterations of the \textit{while} loop of \fptfa (and,  respectively,  $d^{(j)}$ for $d$).
Let us prove by induction on the level $u \in \{1,\cdots,k\}$ the following property $H(u)$:
\begin{subequations}
\begin{empheq}[left={H(u)$=$\empheqlbrace\,}]{align}
&\forall i \in \calI \text{ s.t.  } \dep(i) = u,  \,\, d_{i}^{(2u)} = d_{i}^{(2u-1)} \label{eq:Hindd}\\
&\forall i \in \calI \text{ s.t.  } \dep(i) = u,   \,\, z_{(i,\cdot)}^{(2u+1)} = z_{(i,\cdot)}^{(2u)} \label{eq:Hindz}
\end{empheq}
\end{subequations}

\begin{itemize}
\item \textit{Base step, $u=1$:} Let $i \in  \calI \text{ such that  } \dep(i) = 1$. 
\begin{enumerate}[label=B-\arabic*)]
\item \label{it:B1d} By applying Line~4 of Algorithm~\ref{alg:SyncTFA} at the first loop iteration, we have $d_i^{(1)} = \calD_i(z^{(0)}_{\inc(i)})$. 
The same line for the second loop iteration gives $d_i^{(2)} = \calD_i(z^{(1)}_{\inc(i)})$. 
By definition of the level number, as $\dep(i) = 1$, there are only fresh flows at input of node $i$. 
Hence,  $\inc(i) = \emptyset$, and thus $d_i^{(1)}$ and $d_i^{(2)}$ are independent of $z$.
Therefore,  $d_i^{(2)} = d_i^{(1)}$ and Eq.~\eqref{eq:Hindd} is true.

\item As there are only fresh flows, for each fresh flow $f$ at node $i$, the estimated propagated burstinesses at the output links are 
$$[z_{(i,\cdot)}^{(3)}]_f = [\calY_{(i,\cdot)}(d_{\pred((i,\cdot))}^{(2)})]_f = r_fd_i^{(2)} + b_f$$ 
With ~\ref{it:B1d},  $d_i^{(2)} = d_i^{(1)}$, thus $z_{(i,\cdot)}^{(3)} = z_{(i,\cdot)}^{(2)}$ by definition of $\calY$. 
As a result,  Eq.~\eqref{eq:Hindz} is true,  and $H(1)$ is true.
\end{enumerate}

\item \textit{Induction step:} Assume $H(1),\cdots,H(u-1)$ and let us show $H(u)$. 
Let $i \in \calI$ such that $\dep(i) = u$.
\begin{enumerate}[label=I-\arabic*)]
\item \label{it:I1d} By applying Line~4 of Algorithm~\ref{alg:SyncTFA} at iteration $2u$,  $d^{(2u)}_i = \calD_i(z^{(2u-1)}_{\inc(i)})$.
By definition, $\inc(i)$ is the set of transit edges that are incident to node $i$, thus 
$$\forall j \in \inc(i),  \exists f \text{ such that } i \in \paf(f)$$
With Eq.~\eqref{eq:defprof},  $\dep(j) \leq u-1$.
And by induction assumption, $\forall j \in \inc(i), z^{(2(u-1)+1)}_j = z^{(2(u-1))}_j$, thus $z^{(2u-1)}_j = z^{(2u-2)}_j$.
Therefore $d^{(2u)}_i = \calD_i(z^{(2u-1)}_{\inc(i)}) = \calD_i(z^{(2u-2)}_{\inc(i)}) = d^{(2u-1)}_i$. 
Hence,  Eq.~\eqref{eq:Hindd} is true.
\item Let $f \text{ such that } i \in \paf(f)$, the output burstiness at node $i$ for iteration $2u+1$ for flow $f$ is 
\begin{eqnarray}
[z_{(i,\cdot)}^{(2u+1)}]_{f} &=& \calY_{(i,\cdot)}(d_{\pred((i,\cdot))}^{(2u)}) 
\end{eqnarray}
$\forall j \in \pred((i,\cdot)), \dep(j) \leq u $,  hence with~\ref{it:I1d},  $d_j^{(2u)} = d_j^{(2u-1)}$.
Therefore, 
$$[z_{(i,\cdot)}^{(2u+1)}]_{f} = \calY_{(i,\cdot)}(d_{\pred((i,\cdot))}^{(2u-1)}) = [z_{(i,\cdot)}^{(2u)}]_{f},$$
where the second equality comes from the definition of $\calY$.
Therefore,  $z_{(i,\cdot)}^{(2u+1)} = z_{(i,\cdot)}^{(2u)}$.
Hence,  Eq.~\eqref{eq:Hindz} is true.
\end{enumerate}
\end{itemize}
\end{proof}

\subsection{Equivalence of \synctfa and  \fptfa \,/ \tfa}
We now return to the question of Section~\ref{sec:fptfa}: 
Does the choice of a cut influence the convergence and the value of the bounds computed by \fptfa~? 
To prove that this is not the case, we prove that \synctfa and \fptfa with cut $L$ compute the same thing.
Specifically,  Theorem~\ref{th:fptfa} shows that Algorithm~\ref{alg:SyncTFA} and~\ref{alg:FPTFA} both diverge or both converge; 
and if they both converge,  they obtain the same delay-jitter and burstiness bounds.
This holds for any valid cut, \ie , any cut that leaves the network free of cyclic dependency.
\begin{theorem}
\begin{enumerate}
\item $\lo(\calG)$ bounded $\Leftrightarrow$ $\lo(\calF^L)$ bounded.
\item If $\lo(\calG)$ and $\lo(\calF^L)$ are bounded then 
\begin{enumerate}
\item $\calG$ has a unique fixpoint $(\bar{z},\bar{d})$ and $\calF^L$ has a unique fixpoint $z^*_L$;
\item Let $d^*$ the collection of delay-jitter bounds computed by \fptfa; then $\bar{z}_L=z^*_L$ and $\bar{d}=d^*$.
\end{enumerate}
\end{enumerate}
\label{th:fptfa}
\end{theorem}
\begin{proof}\textbf{Part A.}
In this part, we show that 
\begin{center}
if $(\bar{z},\bar{d})$ is a (finite) fixpoint of $\calG$, then $\calF^L(\bar{z}_L) = \bar{z}_L$.
\end{center}
To prove this implication, we assume that $(\bar{z},\bar{d})$ is a (finite) fixpoint of $\calG$ and we construct $z'_\ell$, $z_\ell$ and $d$ by applying the same construction as in the inner loop of \fptfa, 
which enables us to compute $\calF^L(\bar{z}_L)$. 
We execute specifically  the following algorithm with input value $\bar{z}_L$, \ie, we compute \textsc{FPTFAiter}($\bar{z}_L$).
\begin{algorithmic}[1]
\Function{FPTFAiter}{$z^{0}_L$} 
		\State  $\forall \ell \in L$, $z_{\ell} = z^{0}_\ell$\label{alg:init}
		\For{$k \leftarrow 1 $ to $n$}
		\State $d_{i_k} \gets \calD_{i_k}\lp z_{\inc(i_k)} \rp$ \label{alg:d}
		\State $\forall \ell \in \out(i_k)\setminus L$, $z_{\ell} = \calZ_{i_k} \lp d_{i_k}, z_{\inc(i_k)} \rp$ \label{alg:z}
		\State $\forall \ell \in \out(i_k) \cap L$, $z'_\ell =\calZ_{i_k} \lp d_{i_k}, z_{\inc(i_k)} \rp$ \label{alg:zp}
		\EndFor
		\Output $ (z'_L,z_{\calL \setminus L},d)$     
		\State\Comment $z'_L$ is equal to $\calF^L(z^0_L)$
		\EndOutput
\EndFunction
\end{algorithmic}
We note $z^k_\ell$ the value of $z_\ell$ after the k-th iteration (respectively $z'^{k}_{\ell}$ for $z'_\ell$).
\textsc{FPTFAiter}($\bar{z}_L$) has the following properties:

\begin{eqnarray}
&&\forall \ell \in L,  \forall  k \in \lc 1,...,n\rc, z^k_{\ell} = \bar{z}_{\ell} \label{eq:zkzl} \\
&& \text{If } \ell \in \out(i_{k'}) \setminus L, \forall k \geq k',  z_{\ell}^k = z_{\ell}^{k'} \label{eq:znLrecc} \\
&& \text{If } \ell \in \out(i_{k'}) \cap L,  \forall k \geq k',  z'^{k}_{\ell} = z_{\ell}^{k'} \label{eq:zLrecc}
\end{eqnarray}
Eq.~\eqref{eq:zkzl} holds as in line~\ref{alg:d} to \ref{alg:zp}, $\forall \ell \in L$, $z_\ell$ is not assigned.\\
Eq.~\eqref{eq:znLrecc} -~\eqref{eq:zLrecc} hold as $\forall \ell,  \exists ! k,$ s.t. $\ell \in \out(i_k)$.  
%
By induction on $k \in \lc 1,...,n\rc$,  we now prove the following property~$P_1(k)$~:
\begin{subequations}
\begin{empheq}[left={P_1(k)$=$\empheqlbrace\,}]{align}
&\forall \ell \in \out(i_k) \setminus L,  \forall f \text{ s.t. } i_k \in \pred_f(\ell),  \nonumber\\
&\hspace{2cm}z^k_{\ell ,f} = \bar{z}_{\ell ,f}  \label{eq:zkP} \\
&\forall \ell' \in \out(i_k) \cap L,  \forall f \text{ s.t. } i_k \in \pred_f(\ell), \nonumber\\ 
&\hspace{2cm}z'^{k}_{\ell' ,f} = \bar{z}_{\ell' ,f}  \label{eq:zpkP1}
\end{empheq}
\end{subequations}
%
$\bullet$ \textit{Base step, $k$=$1$:}
\begin{enumerate}[label=B\arabic*)]
\item \label{it0:deadline} By line~\ref{alg:d}, $d_{i_1} = \calD_{i_1}(z_{\inc(i_1)})$.
By Eq.~\eqref{set:i1}, $\inc(i_1) \subset L$.  
Thus,  by line~\ref{alg:init}, $z_{\inc(i_1)} = \bar{z}_{\inc(i_1)}$.

By fixpoint assumption of $\calG$, $\bar{d}_{i_1} = \calD_{i_1}(\bar{z}_{\inc(i_1)})$.

Thus, $d_{i_1}=\bar{d}_{i_1}$.
\item Let $\ell \in \out(i_1) \setminus L$, let $f$ a flow such that $i_1 \in \pred_f(\ell)$.
\begin{enumerate}
\item \label{it0:freshflow} Either $i_1$ is the first hop on the path of $f$ ($f$ is a fresh flow at $i_1$) and by line~\ref{alg:z} and definition of $\calZ$, $z^1_{\ell ,f} =b_f + r_fd_{i_1}$.

As $d_{i_1}=\bar{d}_{i_1}$, $z^1_{\ell ,f} = b_f + r_f\bar{d}_{i_1}$.
And since $\bar{z}$ is a fixpoint of $\calG$ and by definition of $\calY_\ell$,  $\bar{z}_{\ell ,f} = b_f + r_f\bar{d}_{i_1}$.

As a result,  $z^1_{\ell ,f} = \bar{z}_{\ell ,f}$.

\item \label{it0:cutflow} Or $i_1$ is not the first hop on the path of $f$ and $\exists \, \ell'' \in \inc(i_1)$ such that flow $f$ crosses the link $\ell''$.
By line~\ref{alg:z} and definition of $\calZ$, $z^1_{\ell ,f} = z_{\ell'' ,f} + r_fd_{i_1}$.
By Eq.~\eqref{set:i1}, $\inc(i_1) \subset L$, so $\ell'' \in L$, and by Line~\ref{alg:init}, $z_{\ell'' ,f} = \bar{z}_{\ell'' ,f}$.

Hence,  $z^1_{\ell ,f} =  \bar{z}_{\ell'' ,f} + r_f\bar{d}_{i_1}$.

Since $\bar{z}$ is a fixpoint of $\calG$, and by definition of $\calY$,  $\bar{z}_{\ell'' ,f} =  b_f + r_f \sum_{u \in \pred_f(\ell'')} \bar{d}_u$.
By concatenation of paths,  since $\ell'' = (\cdot, i_1)$, and $\ell = (i_1,\cdot)$, $\pred_f(\ell) = \{i_1\} \cup \pred_f(\ell'')$. 
Therefore,  we have 
$$z^1_{\ell ,f} =  b_f + r_f \sum_{u \in \pred_f(\ell'') \cup \{i_1\}} \bar{d}_u.$$

By fixpoint of $\calG$ and definition of $\calY$, 
$$\bar{z}_{\ell ,f} = b_f + r_f \sum_{u \in \pred_f(\ell)} \bar{d}_u.$$

As a result,  $z^1_{\ell ,f}  = \bar{z}_{\ell ,f}$ and Eq.~\eqref{eq:zkP} is true for $k=1$.
\end{enumerate}

\item Let $\ell' \in \out(i_1) \cap L$, let $f$ a flow such that  $i_1\in \pred_f(\ell')$. 
Either $i_1$ is the first hop on the path of $f$ (fresh flow), or it is not  and $\exists \, \ell'' \in \inc(i_1)$ such that flow $f$ crosses the link $\ell''$.

Both cases lead to $z'^{1}_{\ell' ,f} = \bar{z}_{\ell' ,f}$ by using arguments from~\ref{it0:freshflow}) and~\ref{it0:cutflow}).
Eq.~\eqref{eq:zpkP1} is true for $k=1$, and so $P_1(1)$ is true.
\end{enumerate}
$\bullet$ \textit{Induction step:} Assume $P_1(1), \cdots ,P_1(k-1)$, and show $P_1(k)$.
\begin{enumerate}[label=I\arabic*)]
\item \label{itk:deadline} Let $\ell \in \inc(i_k)$:
\begin{enumerate}
\item Either,  $\ell \in \inc(i_k) \cap L$, and by Eq.~\eqref{eq:zkzl}, $z^{k-1}_{\ell ,f} = \bar{z}_{\ell ,f}$.
\item Or,  $\ell \in \inc(i_k) \setminus L$,  so by Eq.~\eqref{set:ik}, $\ell \in \out(i_1) \cup ... \cup \out(i_{k-1})$ and
$\exists k' \in \{1, \cdots , k-1\}$ such that $\ell = (i_{k'},i_k)$.  By induction assumption, $z_{\ell,f}^{k'} = \bar{z}_{\ell,f}$.
By Eq.~\eqref{eq:znLrecc},  $z^{k-1}_{\ell ,f} = z^{k'}_{\ell ,f}$.
Thus $z^{k-1}_{\ell ,f} = \bar{z}_{\ell,f}$.
\end{enumerate}
Then with $\ell \in \inc(i_k)$,  $z^{k-1}_{\inc(i_k)} = \bar{z}_{\inc(i_k)}$, and so $d_{i_k}=\bar{d}_{i_k}$.
%
\item \label{itk:z}  Let $\ell \in \out(i_k) \setminus L$, let $f$ a flow such that $i_k \in \pred_f(\ell)$.
\begin{enumerate}
\item \label{itk:freshflow} Either $i_k$ is the first hop on the path of $f$ (fresh flow) and with same arguments as~\ref{itk:deadline}, and~\ref{it0:freshflow}), $z^k_{\ell ,f} = \bar{z}_{\ell ,f}$.
\item \label{itk:notfreshflow} Or $i_k$ is not the first hop on the path of $f$ and $\exists \, \ell'' \in \inc(i_k)$ such that flow $f$ crosses the link $\ell''$.
\begin{enumerate}[label=b\arabic*)]
\item Either $\ell'' \in L$.
By line~\ref{alg:z}, $z^k_{\ell ,f} = z^{k-1}_{\ell'' ,f} + r_fd_{i_k}$.
Since $\ell'' \in L$, by Eq.~\eqref{eq:zkzl}, $z^{k-1}_{\ell'' ,f} = \bar{z}_{\ell'' ,f}$.
And $d_{i_k}=\bar{d}_{i_k}$, so $z^k_{\ell ,f} =  \bar{z}_{\ell'' ,f} + r_f\bar{d}_{i_k}$.
\ref{it0:cutflow}) arguments lead to $z^k_{\ell ,f} = \bar{z}_{\ell ,f}$.
\item Or $\ell'' \notin L$. By Eq.~\eqref{set:ik} and \ref{itk:deadline} arguments, $z^{k-1}_{\ell'' ,f} = \bar{z}_{\ell'' ,f}$.
By line~\ref{alg:z}, $z^k_{\ell ,f} = z^{k-1}_{\ell'' ,f} + r_fd_{i_k}$, and by arguments from~\ref{it0:cutflow}), $z^k_{\ell ,f} = \bar{z}_{\ell ,f}$.

Therefore,  Eq.~\eqref{eq:zkP} is true for $k$.
\end{enumerate}
\end{enumerate}
\item Let $\ell' \in \out(i_k) \cap L$, let $f$ a flow such that $i_k \in \pred_f(\ell')$. \\ 
Either $i_k$ is the first hop for $f$, or it is not,  and $\exists \, \ell'' \in \inc(i_k)$ such that flow $f$ crosses the link $\ell''$.
As previously,  the analysis is done for cases $\ell'' \in L$, or $\ell'' \notin L$.
In all cases, the arguments in \ref{itk:z} lead to $z'^{k}_{\ell' ,f} = \bar{z}_{\ell' ,f}$. 

Eq.~\eqref{eq:zpkP1} is true for $k$,  hence $P_1(k)$ is true.
\end{enumerate}
As a result,  Eq.~\eqref{eq:zLrecc} gives $\forall \ell \in L,  \exists k',z'^{n}_{\ell} = z^{k'}_{\ell}$,
and by Eq.~\eqref{eq:zkzl}, $z^{k'}_{\ell} = \bar{z}_{\ell}$.
Thus,  $z'_L = \bar{z}_L$.

\textbf{Part B.} Conversely, in this part we show that 
\begin{quote}
If $z^*_L$ is a (finite) fixpoint of $\calF^L$, then we can extend $z^*_\ell$ for all $\ell \in \calL$ and give values to $d^*_{i}$ for all nodes $i$ such that ($z^*,d^*$) is a fixpoint of $\calG$.
\end{quote} 
We thus assume that $z^*_L$ is a (finite) fixpoint of $\calF^L$; by applying \textsc{FPTFAiter}($z^*_L$) = $[z^{'}_{L},z^*_{\calL \setminus L},d^*]$, we extend $z^*_\ell$ for all $\ell \in \calL$ and give values to $d^*_{i}$ for all nodes $i$.
Eq.~\eqref{eq:zpzFPTFA} holds by definition of \fptfa. Eq.~\eqref{eq:zpze} holds since $z^*_L$ is a fixpoint of $\calF^L$,
\begin{eqnarray}
&&\forall \ell \in L, z'_\ell = \left[\calF^L(z^*_L)\right]_\ell \label{eq:zpzFPTFA} \\
&&\forall \ell \in L, z'_\ell = z^*_\ell \label{eq:zpze}
\end{eqnarray}
$\forall f$,  let us prove $P^f_2(h)$ by induction on $h \in [1,n_h]$, with $n_h$ the number of hops of flow $f$ and $i^{h,f}$ its $h^{th}$ hop:
\begin{subequations}
\begin{empheq}[left={P^f_2(h)$=$\empheqlbrace\,}]{align}
&\forall \ell \in \out(i^{h,f}) \setminus L \text{ s.t. } i^{h,f} \in \pred_f(\ell), \nonumber \\
&z^*_{\ell ,f} = b_f + r_f  \sum_{u \in \pred_f(\ell)} d^*_u  \label{eqrec:zkP} \\
&\forall \ell' \in \out(i^{h,f}) \cap L \text{ s.t. } i^{h,f} \in \pred_f(\ell'), \nonumber\\
&z'_{\ell' ,f} = b_f + r_f  \sum_{u \in \pred_f(\ell')} d^*_u \label{eqrec:zpkP1}
\end{empheq}
\end{subequations}
$\bullet$ \textit{Base step, $h$=$1$:} $i^{1,f}$ is the first hop of the flow $f$. 
\begin{enumerate}[label=PB\arabic*)]
\item \label{Pit0:z} Let $\ell \in \out(i^{1,f}) \setminus L$,  by line~\ref{alg:z} and definition of $\calZ$, $z^*_{\ell ,f}=b_f + r_f d^*_{i^{1,f}}$.
As $f$ is a fresh flow,  $\pred_f(\ell) = \{i^{1,f}\}$,  hence Eq.~\eqref{eqrec:zkP} is true.
\item Let $\forall \ell' \in \out(i^{h,f}) \cap L$,  Eq.~\eqref{eqrec:zpkP1} is satisfied with same arguments as~\ref{Pit0:z}.

Therefore,  $P^f_2(1)$ is true.
\end{enumerate}
$\bullet$ \textit{Induction step:} Assume $P^f_2(h-1)$, and show $P^f_2(h)$.
\begin{enumerate}[label=PI\arabic*)]
\item \label{Pitk:notcut} Let $\ell \in \out(i^{h,f}) \setminus L$.
As $h>1$, $i^{h,f}$ is not the first hop of flow $f$, thus $\exists \, \ell'' \in \inc(i^{h,f})$ such that flow $f$ crosses the link $\ell''$.
\begin{enumerate}
\item \label{Pitk:notcutWithcut} Either $\ell'' \in L$,  and by line~\ref{alg:z} and definition of $\calZ$, $z^*_{\ell ,f}=z^*_{\ell'' ,f} + r_f d^*_{i^{h,f}}$.
Since $\ell'' \in L$, by Eq.~\eqref{eq:zpze}, $z'_{\ell'' ,f} = z^*_{\ell'' ,f}$, and $\ell'' \in \out(i^{h-1,f})$, so by induction assumption,  $z^*_{\ell'' ,f} = b_f + r_f  \sum_{u \in \pred_f(\ell'')} d^*_u$.
As $\pred_f(\ell) = \{i^{h,f}\} \cup \pred_f(\ell'')$, Eq.~\eqref{eqrec:zkP} is satisfied.
\item \label{Pitk:notcutnotcut} Or $\ell'' \notin L$.
By line~\ref{alg:z} and definition of $\calZ$, $z^*_{\ell ,f}=z^*_{\ell'' ,f} + r_f d^*_{i^{h,f}}$.
Since $\ell''\in \out(i^{h-1,f})$, then by induction assumption, $z^*_{\ell'' ,f} = b_f + r_f  \sum_{u \in \pred_f(\ell'')} d^*_u$.
As $\pred_f(\ell) = \{i^{h,f}\} \cup \pred_f(\ell'')$, Eq.~\eqref{eqrec:zkP} is satisfied.
\end{enumerate}
\item Let $\ell' \in \out(i^{h,f}) \cap L$, and analyse line~\ref{alg:zp}.
As $h>1$, $i^{h,f}$ is not the first hop of flow $f$, thus $\exists \, \ell'' \in \inc(i^{h,f})$ such that flow $f$ crosses the link $\ell''$.
As in~\ref{Pitk:notcut},  the analysis is done for cases $\ell'' \in L$, or $\ell'' \notin L$.
All these cases lead to $z'_{\ell' ,f} = b_f + r_f  \sum_{u \in \pred_f(\ell')} d^*_u$ by using the same arguments as in~\ref{Pitk:notcutWithcut} and~\ref{Pitk:notcutnotcut}.

Eq.~\eqref{eqrec:zpkP1} is true,  hence $P_1(k)$ is true.
\end{enumerate}
Therefore,  we have 
$$\forall f, \forall \ell \in \calL,  z^*_{\ell ,f} =  b_f + r_f  \sum_{u \in \pred_f(\ell)} d^*_u = \calY_\ell(d^*_{\pred(\ell)}).$$
Also, by definition of $\calD$: 
$$\forall i, d^*_{i} = \calD(z^*_{\inc(i)}).$$

Thus ($z^*,d^*$) is a fixpoint of $\calG$.

\textbf{Part C.} The mapping $\calF^L$ is isotonic and concave by properties of the delay-jitter and burstiness functions.
Indeed,  by the definition in~\cite[Eq.(3)]{thomas2019cyclic}, at each node, the delay-jitter at a node is the minimum of affine functions in input bursts at this node. 
Thus,  the delay-jitter is concave in the input bursts at this node.

By strong recurrence, for all nodes belonging to the feed-forward network (linked with the function $\calF^L$),
the delay-jitter and burstiness functions are concave in the burstinesses at cuts $z_L$, and so $\calF^L$ is concave in $z_L$.
Furthermore,  $\calF^L(0)>0$ as every component in $z'_{\ell}$ is lower bounded by the burstiness at the source of the corresponding flow.

Therefore,  we can apply Corollary~\ref{coro-c1} to $\calF^L$: 
\begin{itemize}
\item Either,  $\lo(\calF^L)$ is unbounded,
\item Or,  $\calF^L$ has a unique fixpoint,  that is also the largest element of $\lo(\calF^L)$. 
\end{itemize}
In the previous section, we show that the same holds for $\calG$. 
It follows from parts A and B that 
\begin{center}
$\calG$ has a fixpoint $\Leftrightarrow$ $\calF^L$ has a fixpoint,
\end{center}
that shows item 1). 

Now,  assume that $\lo(\calG)$ and $\lo(\calF^L)$ are bounded. 
Item 2) similarly follows from parts A and B and from the uniqueness of the fixpoints.
\end{proof} 

\begin{remark}
We can also apply \fptfa with $L=\emptyset$ on a network without cyclic dependencies. 
In such case,  as noticed in Section~\ref{sec:fptfa},  \fptfa and \tfa compute the same bounds. 
Theorem~\ref{th:fptfa} ensures that \synctfa and \fptfa  also compute the same bounds.  
Therefore bounds computed by \synctfa and \tfa are equal.
\end{remark}

\section{Asynchronous TFA}
\label{sec:async-tfa}
We introduce the asynchronous algorithm,  a new algorithm that,  contrary to \synctfa algorithm seen in Section~\ref{sec:sync-tfa}, 
 updates together the delay and burst. 
 Here, we perform asynchronously the TFA updates,  at one or several nodes at a time and in some arbitrary order.

The advantage of \asynctfa is that several nodes can be analysed at the same time,  hence the update of the delays and the burstinesses can be realised for several nodes in parallel. 

Specifically,  at every round $k=1,2,...$, we pick a set of nodes $I_k\subset \calI$;
then,  the TFA update of delay and burstinesses are done simultaneously for every node $i\in I_k$.

We assume

\begin{description}
  \item[\hd] Every node $i$ is visited infinitely often, \ie , $\forall i$ there is an infinite number of rounds $k$ such that $i \in I_k$.
\end{description}

\newcommand\comm[1]{\hfill // #1}
\begin{algorithm}
	\caption{Asynchronous TFA (\asynctfa) \label{alg:AsyncTFA}}
	\begin{algorithmic}[1]
		\State $z_{\ell}\gets 0$, $\forall\ell \in \calL$; 
		\State $d_i\gets 0$, $\forall i \in\calI$; 
		\State $k \gets 0$;
		\While{\texttt{true}}
		\State $k\gets k+1$;
		\State \textbf{parfor }{$i \in I_k$} \textbf{do} \comm{Parallel \textit{for} loop} \label{alg:forparbeg}
		\State \hspace{0.5cm}\textbf{parallelSections do} \label{alg:secparbeg}
		\State \hspace{1cm}\textbf{section1} \label{alg:sec1beg}
		\State \hspace{1.5cm}$d_{i}\gets \calD_{i}\lp z_{\inc(i)} \rp$;\label{eq_alg:delay}
		\State \hspace{1cm}\textbf{end section1}\label{alg:sec1end}
		\State \hspace{1cm}\textbf{section2} \label{alg:sec2beg}
		\State \hspace{1.5cm}$\forall \ell \in \out(i)$, $z'_\ell =\calY_\ell \lp d_{\pred(\ell)} \rp$;\label{eq_alg:burst}
		\State \hspace{1cm}\textbf{end section2} \label{alg:sec2beg}
		\State \hspace{0.5cm}\textbf{end parallelSections} \label{alg:secparend}
		\State \textbf{end parfor} \label{alg:forparend}
		\State $\forall i \in I_k$, $\forall \ell \in \out(i)$, $z_\ell =z'_\ell$;
		\State $print(z,d)$;
		\EndWhile
	\end{algorithmic}
\end{algorithm}
Algorithm~\ref{alg:AsyncTFA} is a parallelized and non deterministic algorithm.  
The \textit{parfor} loop from lines~\ref{alg:forparbeg} to~\ref{alg:forparend} performs a TFA update for all nodes in the subset $I_k$.  
These updates are done in parallel for each element of $I_k$.
In addition, there are parallel sections inside the \textit{parfor} loop.  
Indeed, for each TFA update, each update of $d$ and $z$ are also done in parallel (lines~\ref{eq_alg:delay} and~\ref{eq_alg:burst}). 
$d$ and $z$ are shared variables, so the most recent values of $d$ and $z$ are used to compute lines~\ref{eq_alg:delay} and~\ref{eq_alg:burst}.  
At round~$k$, burst update at line~\ref{eq_alg:burst} can be done with a new value of $d$ updated in round~$k$ or with a former value of~$d$ from round~$k-1$.
\begin{theorem}
The sequence of $(z,d)$ at each loop of \asynctfa is (widesense) increasing.
If $\lo(\calG)$ is bounded then the \asynctfa sequence converges to the largest element $(\bar{z}, \bar{d})$ of $\lo(\calG)$.
Else it does not converge.
\label{th:theo-async}
\end{theorem}
\begin{proof}
As we start at~$0$ and all operators are isotonic, by induction, the sequence of $(z,d)$ in Algorithm~\ref{alg:AsyncTFA} is (widesense) increasing.

Assume that $\lo(\calG)$ is bounded, by Theorem~\ref{theo-tarski},  it has a larger element $(\bar{z}, \bar{d})$ which is also a fixpoint of $\calG$. Let $(z^k,d^k)$ be the values printed at the end of iteration $k$. We have $(z^0,d^0)=(0,0)\leq (\bar{z}, \bar{d})$, hence, by monotonicity of $\calD_i$ and $\calY_{\ell}$ in lines 7 and 8, $(z^1,d^1)\leq \calG(\bar{z}, \bar{d})=(\bar{z}, \bar{d})$. By induction it comes that $(z^k,d^k)\leq (\bar{z}, \bar{d})$ for all $k\geq0$. As a result,  the sequence is bounded and converges.
By~\hd, every node is visited infinitely often,  hence Line~\ref{eq_alg:delay} and~\ref{eq_alg:burst} are executed infinitely often.
By continuity of $\calD$ and $\calY$, the limit is a fixpoint of $\calG$.
We have shown in the proof of Theorem~\ref{th:lowGbounded} that $\calG$ satisfies the hypotheses of Theorem~\ref{theo-fpconc}, thus $\calG$ has a unique fixpoint. This shows the second statement.

Conversely, if the sequence $(z^k,d^k)$ of \asynctfa converges,
then by continuity of $\calD$ and $\calY$ and by definition of $\calG$,
the limit is a fixpoint of $\calG$.
By Theorem~\ref{theo-fpconc}, this implies that $\lo(\calG)$ is bounded.
By contraposition, $\lo(\calG)$ unbounded implies the non-convergence of the \asynctfa sequence.
\end{proof}

It follows that \asynctfa produces the same end-results as \synctfa, \fptfa/\tfa. It should 
converge more quickly than \synctfa,
as it always uses the most recently available values.

\asyncalttfa is a special case of \asynctfa where all nodes are visited simultaneously at every round, i.e. , $I_k=\calI$ for every $k$,  
and where for all $i \in I_k$, some workers execute line~\ref{eq_alg:delay} in parallel before some workers execute line~\ref{eq_alg:burst} in parallel. 
It corresponds to the case where each parallel section~$1$ and~$2$ are done sequentially one after the other for all nodes at round~$k$.
As a result,  \asyncalttfa computes delay-jitter bounds at every node, assuming propagated burstinesses are known,
then it updates the propagated burstinesses, until it finds a fixpoint or diverges.  
\hd holds,  hence all the statements in Theorem~\ref{th:theo-async} hold for \asyncalttfa. 
In particular,  \asyncalttfa produces the same end-results as \synctfa, \fptfa/\tfa.
\begin{algorithm}	\label{alg:AsyncAltTFA}
	\caption{Alternating TFA (\asyncalttfa)}
	\begin{algorithmic}[1]
		\State $z_{\ell}\gets 0$, $\forall\ell \in \calL$; 
		\State $d_{i}\gets 0$, $\forall i \in \calI$;
		\While{\texttt{true}}
		\State $d\gets \calD(z)$;
		\State $z\gets \calY(d)$;
		\State $print(z,d)$;
		\EndWhile
	\end{algorithmic}
\end{algorithm}

\asynctfa algorithms are also valid for networks without cyclic dependencies and becomes stationary in a finite number of iteration.  
Theorem~\ref{th:level-async} specifies the maximal number of iteration steps before stability for \asyncalttfa.
\begin{theorem}
In a network without cyclic dependencies, \asynctfa converges in a finite number of iterations. 
The number of iterations depends on the organisation of the subsets of nodes on Line 6 in Algorithm~\ref{alg:AsyncTFA}. 
In the specific case of \asyncalttfa with a network of $k$~levels, \asyncalttfa is stationary in at most $k$ steps.
\label{th:level-async}
\end{theorem}
\begin{proof}
\asynctfa converges in a finite number of iterations,  due to Assumption~\hd.
The proof of stationary of \asyncalttfa in at most $k$ steps is similar to the proof of Theorem~\ref{th:level}.
\end{proof}

\section{Background: Fixpoints and Maximal Elements}
\label{sec:fixpointsMaxElem}
We recall that for \synctfa and \asynctfa algorithms,  worst-case delay and burstiness bounds in a network (respectively,  $d(t)$ and $z(t)$) belong to the specific set $\lo(\calG) = \lc (z,d), \text {such that } (z,d)\leq \calG(z,d)\rc$, where $\calG$ corresponds to the function that computes burstinesses and delay at each loop of the algorithms.
Bounds belong to a similar set for \fptfa algorithm with the $\calF^L$ function.
The problem here is to find an upper bound on $\lo(\calG)$ (respectively, $\lo(\calF^L)$ for \fptfa).
In this section, we introduce two theorems:
\begin{itemize}
\item Theorem~\ref{theo-tarski} is introduced to prove that our algorithms are valid in the sense that when they converge they find valid worst-case bounds. 
\item Theorem~\ref{theo-fpconc} proves that the solution is unique. 
This is used to show the equivalence of all algorithms. 
\end{itemize}

For every $b \in \Reals^n$, define the set $S(b)$ by
$$S(b) \eqdef \{z \in \Reals^n, z \geq b\},$$
where comparison is coordinatewise. For a function $F: S(b) \to S(b)$, 
we say that $z$ is a fixpoint of $F$ if $z \in S(b)$ and $F(z)=z$, \ie, we consider only finite fixpoints.

We say that $F$ is isotonic if and only if
$$\forall z,z'\in S(b),\; z\leq z' \Rightarrow F(z)\leq F(z').$$

\begin{theorem} \label{theo-tarski}
Let $F: S(b) \to S(b)$ be isotonic.
If the set $\lo(F)\eqdef \lc z\in S(b), z\leq F(z)\rc$ is bounded,
it has a largest element $\bar{z}$ and $F$ has at least one fixpoint.
$\bar{z}$ is also the largest fixpoint of $F$.
\end{theorem}
Theorem~\ref{theo-tarski} is a variant of the Knaster-Tarski theorem~\cite{tarski1955}.
In Remark~1, we present an example that highlights the differences between Theorem~\ref{theo-tarski} and Knaster-Tarski theorem.
\begin{proof}
This proof is based on the proof of the Knaster-Tarski theorem~\cite{tarski1955}.

Let us assume that the set $\lo(F)\eqdef \lc z\in S(b), z\leq F(z)\rc$ is bounded.
We introduce $\bar{z}$ by: 
\begin{eqnarray}
\bar{z} \eqdef \sup(\lo(F)). \label{eq:supzbar}
\end{eqnarray}
Since $\lo(F)$ is bounded, $\bar{z}$ is finite.
$\forall z\in\lo(F)$, $z\leq \bar{z}$.
Since $F$ is isotonic, $F(z)\leq F(\bar{z})$ and since $z\leq F(z)$, it follows that $z\leq F(\bar{z})$.
Thus, $F(\bar{z})$ is an upper bound of $\lo(F)$, hence $F(\bar{z})\geq \sup(\lo(F))=\bar{z}$.
This shows that $\bar{z}\in \lo(F)$, i.e. , $\lo(F)$ has a largest element, $\bar{z}$.

Second, $F(F(\bar{z}))\geq F(\bar{z})$, thus $F(\bar{z})\in \lo(F)$. 
With Eq.~\eqref{eq:supzbar} and previous inequality, we have $F(\bar{z})\leq \bar{z}$. 
Therefore, we conclude that $F(\bar{z})=\bar{z}$.
This shows that $F$ has a fixpoint $\bar{z}$.
Any other fixpoint is in $\lo(F)$ hence is lower than $\bar{z}$.
\end{proof}

\begin{remark}
The usual form of the Knaster-Tarski theorem claims that, if $L$ is a complete lattice and $\phi:L \rightarrow L$ is isotone, then the least upper bound of $\{x\in L \mid x \leq \phi(x)\}$ is a fixpoint of $\phi$. 
It is tempting to apply the Knaster-Tarski theorem by allowing $+\infty$ in the domain of definition of $F$ and in its values, in order to obtain a substitute for \thref{theo-tarski}; however, as we show next, this does not work.

Indeed applying the Knaster-Tarski theorem to an isotone function $\bar{F}: [b;+\infty]^n\to [b;+\infty]^n$ gives that $\bar{F}$ has a fixpoint, 
the set $\lc z\in [b;+\infty]^n, z\leq \bar{F}(z)\rc$ has a largest element, and this largest element is also the largest fixpoint of $\bar{F}$. 
However, as the following example shows, this does not provide the largest \emph{finite} solution of $z\leq \bar{F}(z)$.

\begin{itemize}
\item Consider $n=1$ and $F$ defined by: 
\begin{eqnarray}
 F: [b;+\infty) & \to & [b;+\infty) \label{eq:defF1} \\
 z & \mapsto & \frac12z+\frac12 \label{eq:defF2}
\end{eqnarray} 
It can be extended to an isotone function $\bar{F}: [b;+\infty]\to [b;+\infty]$ by setting $\bar{F}(+\infty)=+\infty$. 

The set $\lc z\in [b;+\infty]^n, z\leq \bar{F}(z)\rc$ is equal to $[0;1] \cup \lc +\infty\rc$. 
Here $F$ has one fixpoint $z=1$ and $\bar{F}$ has two fixpoints: $1$~and~$+\infty$.

\item Also,  consider $F'$ defined by :
\begin{eqnarray}
 F': [b;+\infty) & \to & [b;+\infty) \label{eq:defFp1} \\
 z & \mapsto & z+1 \label{eq:defFp2}
\end{eqnarray} 
$F'$ can also be extended to $\bar{F}': [b;+\infty]\to [b;+\infty]$ by setting $\bar{F}'(+\infty)=+\infty$. 

The set $\lc z\in [b;+\infty]^n, z\leq \bar{F}'(z)\rc$ is equal to $[b;+\infty]$. 
Here,  $F'$ has no fixpoint and $\bar{F}'$ has one fixpoint: $+\infty$.
\end{itemize}
In both cases, the Knaster-Tarski theorem obtains the same result,  \ie, that $+\infty$ is the largest fixpoint of $\bar{F}$ and the largest solution of $z\leq \bar{F}(z)$, which is not helpful. 
In contrast, our results obtain that either $\lc z\in [b;+\infty)^n, z\leq F(z)\rc$ is unbounded and $F$ has no fixpoint (as in the latter case), 
or $F$ has a unique fixpoint (as in the former case) that is also the largest (finite) solution of $z\leq F(z)$.
\end{remark}

Theorem~\ref{theo-fpconc} enables us to show the unicity of the fixpoint, by using additional properties of $\calG$ and $\calF^L$. 
\begin{theorem}
Consider an isotonic function $F: S(b) \to S(b)$ and assume in addition that $F$ is concave and $F(b)>b$.
If $F$ has a fixpoint $z^*$, then
\begin{enumerate}
\item $\lo(F)$ is bounded.
\item The largest element of $\lo(F)$ is $z^*$.
\end{enumerate}
\label{theo-fpconc}
\end{theorem}
\begin{proof}
The proof is inspired by the proof of Lemma~6 in~\cite[Section 6.2]{bouillard2020trade}.

Assume that $F$ has a fixpoint $z^*\in S(b)$. We have $z^*\geq b$ hence $z^*_j=F_j(z^*)\geq F_j(b)>b_j$ for all $j$. 

Now fix some arbitrary $z\in \lo(F)$. We prove by contradiction that $z\leq z^*$.
Assume this does not hold, \ie, there exists some $j$ such that $z^*_j<z_j$ and thus
\begin{eqnarray*}
 0< z^*_j-b_j <z_j - b_j
\end{eqnarray*}  
and 
\begin{eqnarray*}
 \frac{z_j - b_j}{z^*_j-b_j}>1
\end{eqnarray*}  

%
Let $r$ be such that $\frac{z_r-b_r}{z^*_r-b_r}$ is maximum and $\gamma=\frac{z_r-b_r}{z^*_r-b_r}>1$.
It follows that $z_i-b\leq \gamma (z_i^*-b_i)$ for all $i$.
Now,  let $z'=\frac{1}{\gamma}(z-b)+b$.
Thus $z'\leq z^*$ and $z'_r=z_r^*$.
By the former inequality and the isotonicity of $F$,
\begin{equation}
F_r(z') \leq F_r(z^*)=z_r^*
\label{eq-kjnsda}
\end{equation}
where the last equality is because $z^*$ is a fixpoint.  \\
Now $F_r$ is concave, hence
\begin{eqnarray*}
F_r(z')&\geq & \lp 1-\frac{1}{\gamma}\rp F_r(b)+\frac{1}{\gamma}F_r(z)\\
&\geq & \lp 1-\frac{1}{\gamma}\rp F_r(b)+\frac{1}{\gamma}z_r
\end{eqnarray*}
where the latter inequality is because $F(z)\geq z$.\\
Thus,  as $F_r(b)>b_r$ by hypothesis,
\begin{eqnarray*}
F_r(z') &\geq&  \lp 1-\frac{1}{\gamma}\rp F_r(b)+ z'_r \\
&>&  z'_r \\
&=& z^*_r
\end{eqnarray*}
This contradicts Equation~\eqref{eq-kjnsda}. 
Therefore, we have $z\leq z^*$.  
This shows that $\lo(F)$ is bounded by $z^*$ and since $z^*\in\lo(F)$, this shows item 2).
\end{proof}
We can notice that the assumptions in the theorem are satisfied by the functions $\calG$ and $\calF^L$ (with $b=0$).
\begin{corollary}
Under the conditions of Theorem~\ref{theo-fpconc}, one of the following mutually exclusive conditions must hold:
\begin{enumerate}
\item $F$ has a unique fixpoint, $\lo(F)$ is bounded and the fixpoint of $F$ is the largest element of $\lo(F)$;
\item or $F$ has no fixpoint and $\lo(F)$ is unbounded.
\end{enumerate}
\label{coro-c1}
\end{corollary}
In particular for a function $F$ that satisfies the conditions of Theorem~\ref{theo-fpconc}, whenever a sequence converges to a fixpoint of $F$, this proves that $\lo(F)$ is bounded and the limit point must be the largest element of $\lo(F)$.
This is a key argument that allowed us to prove the equivalence of all algorithms in this paper.

\section{Numerical Illustration}
\label{sec:simu}

\begin{figure}[H]
\centering
\includegraphics[width=2.5in]{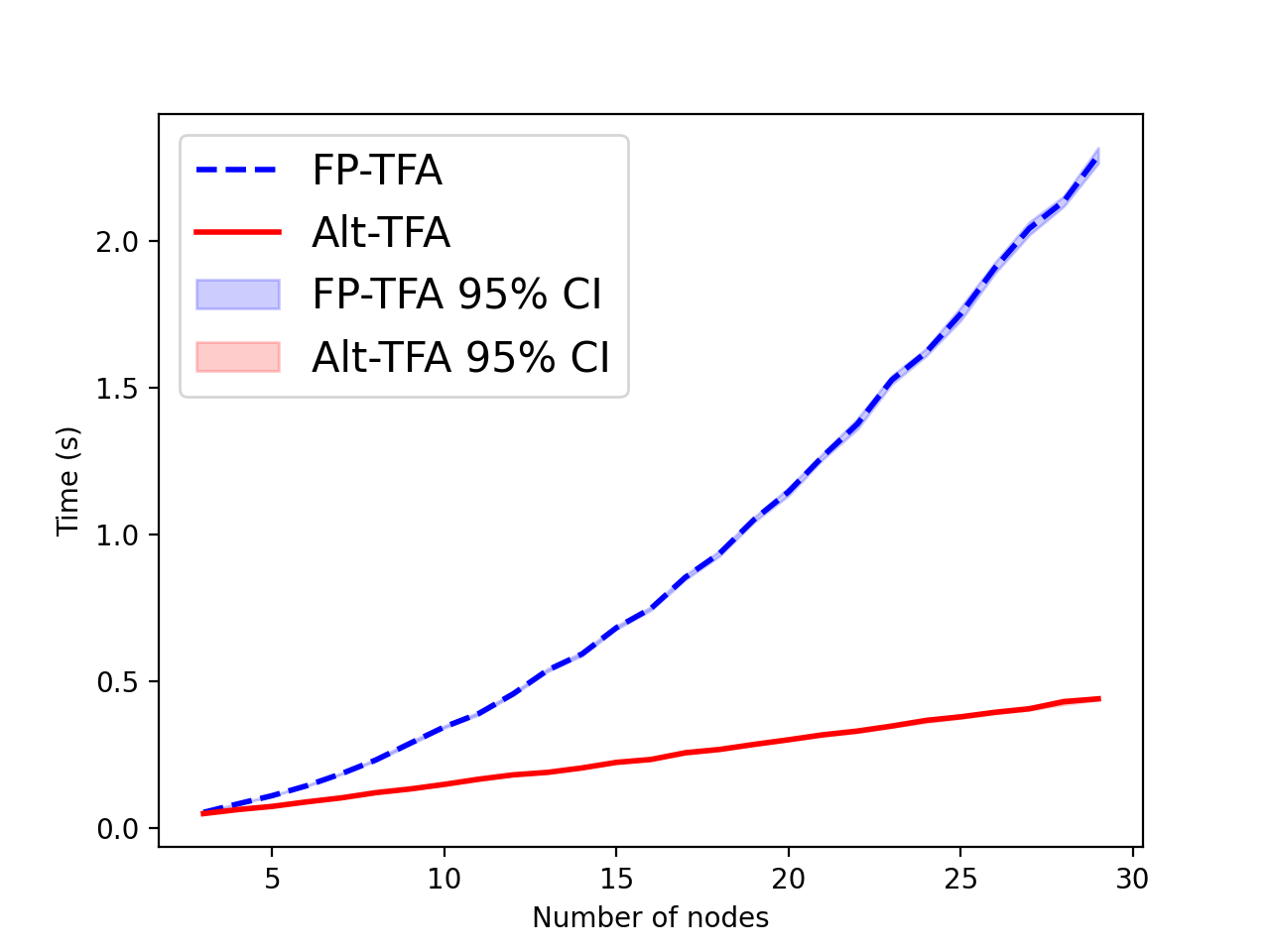}
\caption{Execution time on the ring network,  for \fptfa and \asyncalttfa algorithms and their respective confidence intervals at 95\% (CI).}
\label{fig:simu}
\end{figure}

Figure~\ref{fig:simu} represents the execution times needed to compute the end-to-end delay-jitter for \fptfa and the \asyncalttfa algorithms presented in this article for a flow that crosses~$n-1$ different servers on a ring network of~$n$ servers.  
Every server has a rate-latency service curve with rate $R = 10^7 \texttt{ bits/s}$) and latency $T = 0.001\texttt{s}$. 
There are $n-1$ other flows, each of them starts from a different node and crosses $n-1$~nodes.
They all have the same rate and same input burstinesses ($r =  0.7R/n \texttt{ bits/s}, b = 1000 \texttt{ bits}$). 
The execution time in Figure~\ref{fig:simu} is the mean on $10000$ simulations for each ring size. 
99\% confidence intervals are less than $0.3$\%.
As analysed in Section~\ref{sec:async-tfa}, both algorithms compute the same end-to-end delay-jitter bounds, and the asynchronous algorithms converges faster (see Figure~\ref{fig:simu}).

\section{Conclusion}

In this paper, we have presented new formulations of the TFA and FPTFA algorithms.  
They do not use any cut and work whether the network has cyclic dependencies or not.

These new formulations show us that TFA methods from the literature all produce the same results,  in terms of convergence, and compute the same delay and burstiness bounds.
Furthermore,  the asynchronous version,  presented in Section~\ref{sec:async-tfa},  and its simulation show a gain in computing time.

%
\IEEEpeerreviewmaketitle

\ifCLASSOPTIONcaptionsoff
  \newpage
\fi



%

%
%

\bibliographystyle{IEEEtran}
\vspace{-0.05in}
\bibliography{IEEEabrv,ref}

%
%
%
%
%
%
%
%
\end{document}